\documentclass[12pt]{iopart}

\usepackage{epsfig,bbm,bm,amssymb} 

\def\hf{\frac{1}{2}} 
\def\pq{$\left(p,q\right)\;$} 
\def\dx{\Delta x} 
 
\begin{document} 
          
\title[\pq bound states]{Numerical experiments with $p$ F- and $q$ D-strings: the 
formation of \pq bound states} 
 
\author{Arttu Rajantie$^1$, Mairi Sakellariadou$^2$, Horace Stoica$^1$} 
\address{$^1$ Blackett Laboratory, 
Imperial College London, 
SW7 2AZ U.K.}
\eads{\mailto{a.rajantie@imperial.ac.uk},\;\mailto{f.stoica@imperial.ac.uk}}
\address{$^2$ Department of Physics,  
King's College London, 
London WC2R 2LS, U.K.} 
\ead{Mairi.Sakellariadou@kcl.ac.uk}

\date{\today} 

\begin{abstract}
  We investigate the behaviour of \pq string networks,
  focusing on two aspects: (1) modelling more realistic \pq string
  networks than the $Z_{N}$ networks used so far and (2) investigating
  the effect of long-range interactions on the evolution of the
  network. We model the network with no long-range interactions using
  two sets of fields, complex scalars coupled to gauge fields, with a
  potential chosen such that the two types of strings will form bound
  states.  This way we can model junctions of 3 strings with different
  tension; in $Z_{N}$ models used so far in simulations all the
  strings have identical tensions.  In order to introduce long-range
  interactions we also study a network in which one of the scalars
  forms global strings.
  
  We observe that in the absence of long-range interactions the
  formation of bound states has a significant influence on the
  evolution of the network. When long-range interactions are turned on
  the bound states are short-lived and have a minimal effect on the
  network evolution.  
\end{abstract}
\submitto{JCAP}

\section{Introduction} 
Recent observational data~\cite{Spergel:2006hy} strongly support the
inflationary paradigm~\cite{Guth:1980zm} as a successful solution to
the shortcomings of the standard hot big bang model and the origin of
density fluctuations, leading to the observed large-scale
structure. Nevertheless, despite its success, inflation still remains
a paradigm in search of a model. Successful inflationary models should
thus be motivated by some fundamental physics. Having this in mind,
inflation has been studied extensively in the framework of
supersymmetric grand unified theories, where the end of an
inflationary era is generically accompanied by cosmic string
formation~\cite{Jeannerot:2003qv}. In addition, inflation faces some
issues regarding its onset under generic initial conditions. The
studies on the probability of the onset of
inflation~\cite{Piran:1986dh} indicate that it should take place in
the deep quantum gravity regime. A successful inflationary model could
thus be obtained in the process of brane interactions, in the
framework of brane cosmology within string
theory~\cite{Kallosh:2007ig}.
 
String theory, a consistent theory of quantum gravity, which is 
assumed to be relevant during the early stages of our universe, leads 
to a number of insights into the physics of inflation. One crucial 
issue in cosmological models inspired by string theory is that 
compactification to four space-time dimensions leads to a number of 
scalar fields and moduli. On the one hand, moduli could play the 
r\^ole of the inflaton field, provided they do not roll quickly. On 
the other hand though, runaway moduli would destroy any consistent 
cosmological model. Thus, one requires moduli stabilisation, as 
achieved for instance within the KKLT scenario~\cite{Kachru:2003aw}. 
Among the various proposed brane inflation 
models~\cite{Quevedo:2002xw}, the KKLMMT 
model~\cite{Kachru:2003sx} is a brane-anti-brane annihilation scenario 
in the warped geometry. In this model, the inflaton field is 
associated with the relative position of branes in the compactified 
space, and all moduli are stabilised at the exit from inflation. 
Brane annihilation releases energy which can then reheat the universe, 
thus the radiation-dominated phase of the standard hot big bang 
cosmology can take place. 
 
Consider a Type IIB string theory in 10 space-time dimensions. Brane 
annihilations allow the survival only of three-dimensional 
branes~\cite{Durrer:2005nz}, one of which plays the r\^ole of our 
universe, with the copious production of fundamental (F-strings) and 
Dirichlet D1-branes (D-strings). Such strings are of cosmological size 
and they could play the r\^ole of cosmic 
strings~\cite{Sarangi:2002yt}; they are referred to in the literature as 
cosmic superstrings~\cite{Polchinski:2004hb}.  Individually, the F- 
and D-strings are $\frac{1}{2}$-BPS objects, which however break a 
different half of the supersymmetry each.  In a number of successful 
brane-inflation models, as for example in the KKLMMT model, a spectrum 
of \pq strings, bound states of $p$ F- and $q$ D-strings, with reduced 
tensions was found. The bound \pq states have tension 
\begin{equation} 
\label{pq_tension} 
\mu_{(p,q)}=\mu_{\rm F}\sqrt{p^2+q^2/g_{\rm s}^2}~, 
\end{equation} 
where $\mu_{\rm F}$ denotes the effective fundamental string tension 
after compactification and $g_{\rm s}$ stands for the string 
coupling. The \pq bound states are still $\frac{1}{2}$-BPS objects. 
The presence of stable bound states implies the existence of junctions, 
where two different types of string meet at a point and form a bound 
state leading away from that point.   
 
Traditionally, cosmic strings have been assumed to share the 
characteristics of type-II Nielsen-Olesen (NO) vortices~\cite{Nielsen:1973cs} 
in the Abelian Higgs model. The main differences between these ordinary 
cosmic strings and cosmic superstrings are the following: (i) the 
intercommutation probability for 
ordinary strings is equal to 1, whereas it is smaller (often 
much smaller) than 1 in the case of superstrings~\cite{Jackson:2004zg}; (ii) 
ordinary string networks consist 
of (sub-horizon sized) loops and (super-horizon sized) long strings, 
whereas cosmic superstring networks have also junctions 
at which three string segments meet;
(iii) all strings in an ordinary
string network have the same tension, whereas there is a whole range 
of tensions for superstrings. The last two of these features are 
shared with type-I vortices in the Abelian Higgs model~\cite{Donaire:2005qm}, 
but in contrast with them, cosmic superstrings have two integer-valued 
charges $p$ and $q$.

These differences are expected to imply a different evolution of a
cosmic superstring network as compared to that of ordinary strings. In
particular, a question which has early been addressed is whether such
a network will eventually reach scaling, or whether it will
freeze~\cite{Sen:1997xi}, leading to predictions inconsistent with our
observed universe.  The reader can find in the literature a number of
numerical
experiments~\cite{Sakellariadou:2004wq,Avgoustidis:2004zt,Tye:2005fn,Copeland:2005cy,Saffin:2005cs,Avgoustidis:2005nv,Hindmarsh:2006qn,Avgoustidis:2007aa}
with cosmic superstrings, each of them at a different level of
approximation, as well as analytical studies~\cite{Copeland:2006eh}.
These distinct features are expected to lead to different
observational and cosmological consequences.
 
The aim of our work is to build a simple field model of \pq bound states,
in analogy with the Abelian Higgs model used to investigate 
the properties of ordinary cosmic string networks, and to study its 
properties using lattice simulations. We are mainly
interested in the overall characteristics of the network, and therefore we
focus on the total energy of the network, rather than the dynamics of 
individual strings.  In Section 2, we briefly describe our model.  In 
Section 3, we discuss the \pq string spectrum. In Section 4, we 
describe our numerical approach to studying the \pq string spectrum. 
In Section 5, we present our results from numerical experiments with 
small and large simulations. We round up with our conclusions in 
Section 6. 
 
\section{The model} 
 
We want to build a model of \pq strings which captures the main 
features of the string theory model and is amenable to study via 
lattice simulations. Therefore, we need a system which features: 
\begin{itemize} 
\item Two different species of cosmic strings. We realise that by 
  including two sets of fields of the Abelian Higgs model. 
\item The formation of bound states. We realise that 
  by introducing a coupling of the scalar fields via the potential.  
\item One non-BPS species of cosmic string.  Such strings have
long-range interactions (regardless of their orientation) and this can
be realised by having the second type of string be the topological
defect of a scalar field with a global U(1) symmetry.
\end{itemize} 
 
In the case where both species of cosmic strings are BPS, the action  
reads: 
 \begin{eqnarray} 
\label{equ:action}
 \mathcal{S} &=& \int {\rm d}^{3}x{\rm d}t \biggl[\biggr. 
-\frac{1}{4}F^{2} -\hf \left(D_\mu\phi\right) \left(D^{\mu}\phi\right)^{*}
-\frac{\lambda_{1}}{4}\left(\phi\phi^{*}-\eta_{1}^{2}\right)^{2} 
\nonumber \\ 
&& -\frac{1}{4}H^{2} -\hf \left(D_\mu\phi\right) \left(D^{\mu}\phi\right)^{*}
-\frac{\lambda_{2}}{4}\phi\phi^{*} 
\left(\chi\chi^{*}-\eta_{2}^{2}\right)^{2} 
\biggl.\biggr],  \end{eqnarray}  
where $\phi$ and $\chi$ are two complex scalar fields, and we have used a compact notation for the covariant derivative $D_\mu$, so that 
\begin{eqnarray}
D_\mu\phi&=&\partial_\mu\phi-ie_1A_\mu\phi,\nonumber\\
D_\mu\chi&=&\partial_\mu\chi-ie_2C_\mu\chi.
\end{eqnarray}
In order to avoid
confusion we will refer to the $\phi$ field as the ``Higgs'' and to
the $\chi$ field as the ``axion'', even though both fields are
Higgs-like.  The scalars are coupled to two different U(1) gauge
fields $A_\mu$ and $C_\mu$, with coupling constants $e_1$ and $e_2$
and field strength tensors $F_{\mu\nu}=\partial_\mu A_\nu
-\partial_\nu A_\mu$ and $H_{\mu\nu}=\partial_\mu C_\nu -\partial_\nu
C_\mu$, respectively.  The scalar potentials are parameterised by the
positive constants $\lambda_1, \lambda_2,\eta_1, \eta_2$.
 In the case of a non-BPS species of string, we remove the second 
gauge field by setting $e_2=0$.

The classical equations of motion for the fields follow from the action (\ref{equ:action}),
\begin{eqnarray}
\label{equ:eom}
\partial_\mu F^{\mu\nu}&=&2e_1{\rm Im}\phi^* D^\nu\phi,\nonumber\\
\partial_\mu H^{\mu\nu}&=&2e_2{\rm Im}\chi^* D^\nu\chi,\nonumber\\
D_{\mu} D^\mu \phi &=& -2\lambda_1 \left(\phi^*\phi -\eta_1^2\right)\phi\nonumber\\
D_{\mu} D^\mu \chi &=& -2\lambda_2 \phi^*\phi\left(\chi^*\chi -\eta_2^2\right)\chi.
\end{eqnarray}

Let us look at the potential terms for the two scalar fields. The 
overall value of the potential for the $\chi$ field depends on the 
value of the $\phi$ field at that location. If $\phi$ is in the 
vacuum, $\phi\phi^{*} = \eta_{1}^{2}$, the potential has the usual 
Mexican hat form, while at the core of a $\phi$ vortex we have 
$\phi\phi^{*} = 0$. If a $\chi$ vortex passes through the core of a 
$\phi$ vortex, the potential energy of the $\chi$ field at the core of 
the vortex is (almost) eliminated, so the energy of the vortex is 
reduced. It is therefore favourable for the two vortices to form a 
bound state. 
 
\begin{figure}[htbp] 
  \begin{center} 
    \includegraphics[width=0.35\textwidth,angle=0]{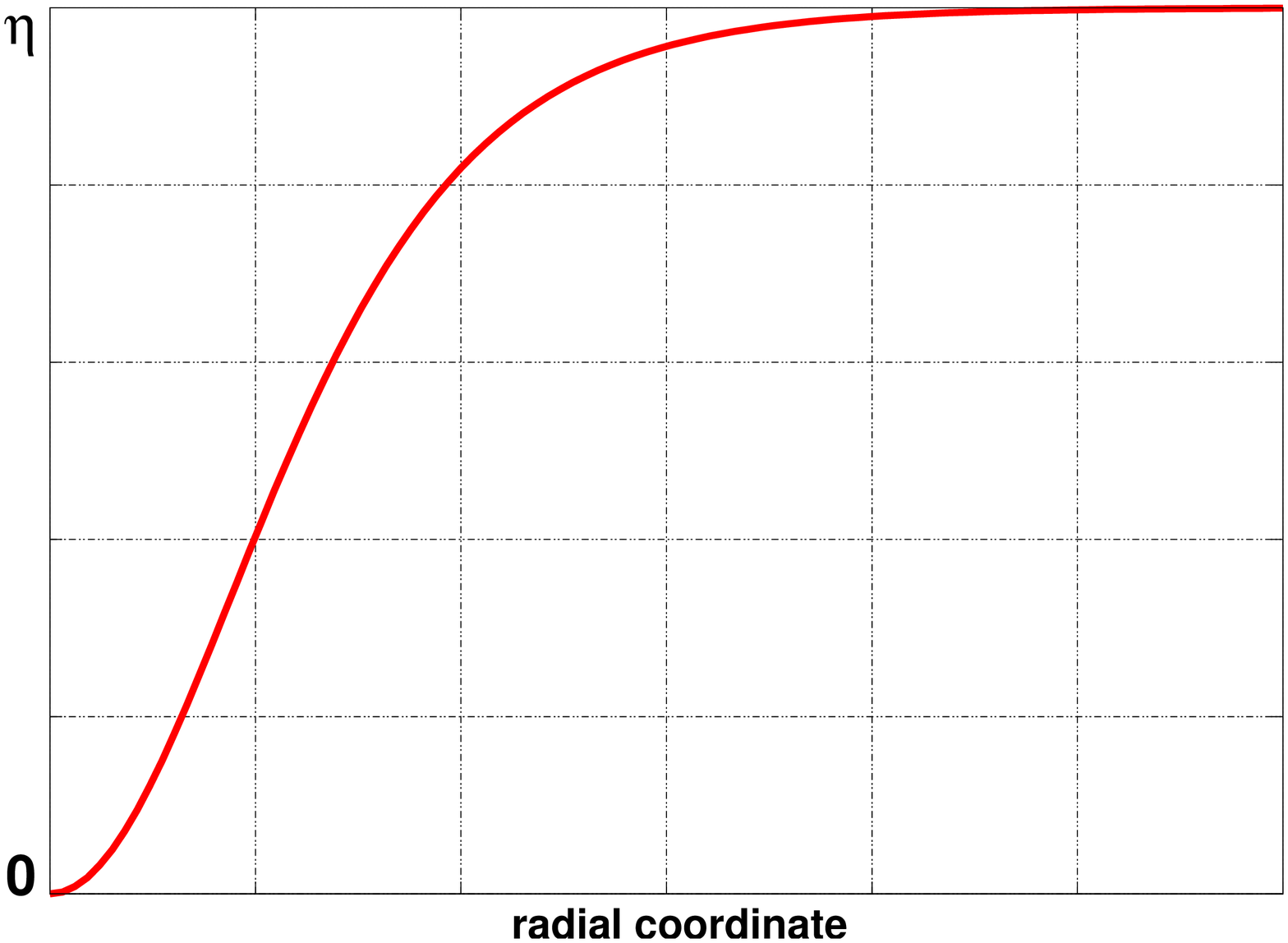} 
    \vspace {18pt} 
    \includegraphics[width=0.4\textwidth,angle=0]{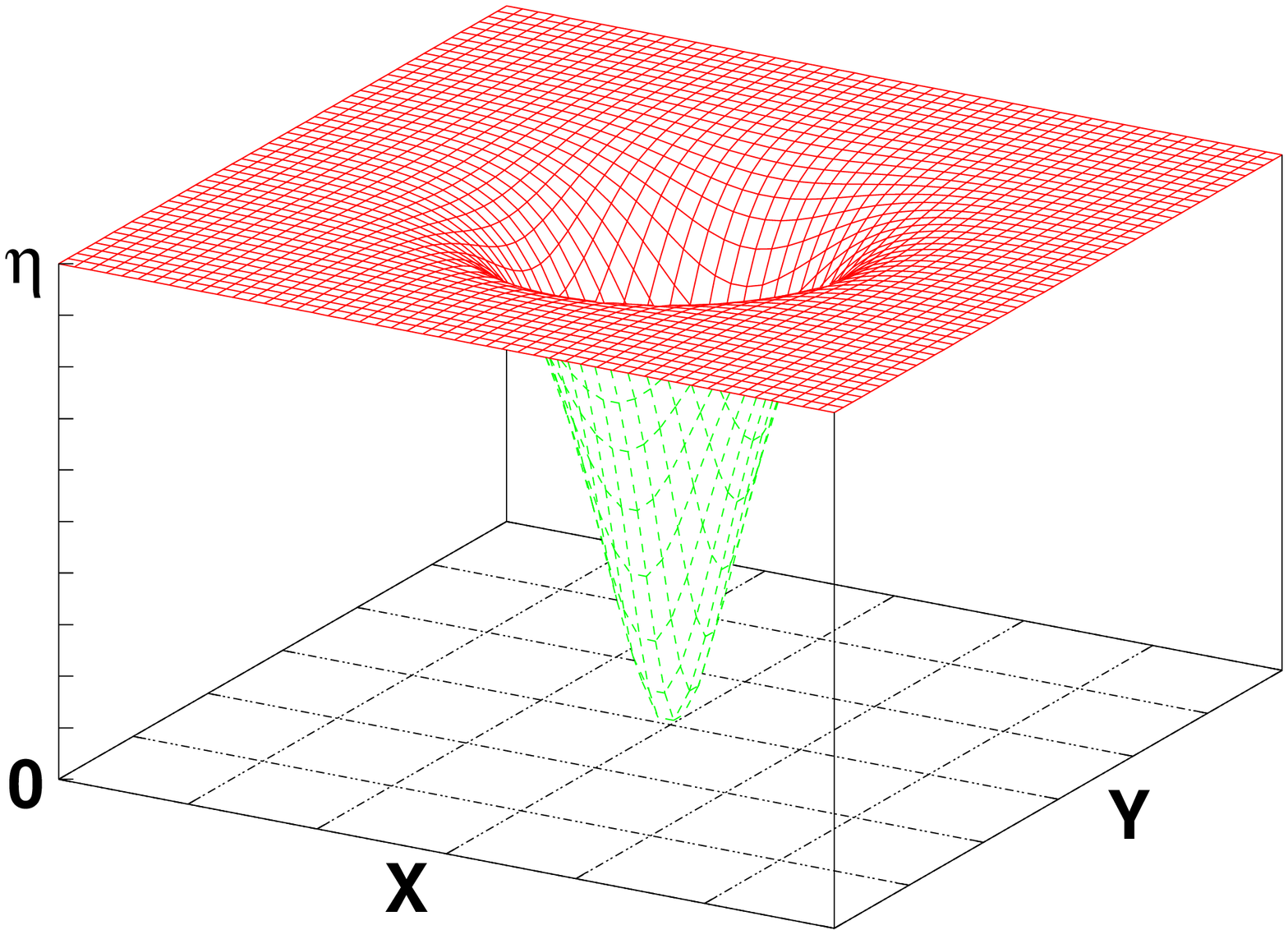} 
  \end{center} 
  \caption{Left panel: Radial field profile for the Higgs field. 
    Right panel: Effective potential as seen by a very thin axion  
    string.   
    We can use the radial field profile of the Higgs field 
    to determine a spatial potential for the axion string. 
    The location corresponding to minimum energy for the axion  
    string is the core of the Higgs string. 
\label{bound_state_potential}} 
\end{figure}

It is easier to illustrate how this works if we look at an extreme 
situation, where the thickness of the axion string is much smaller 
than that of the Higgs string.  Let us consider a Higgs string located 
at the origin of the $x-y$ plane and oriented along the $z$ 
direction. We can use the radial profile of the Higgs field, 
$\phi\left(r\right)$, to determine the radial profile of the function, 
$\phi\phi^\star$, which multiplies the axion field potential. Since the 
axion field potential energy is non-zero only in a very small region, 
the function 
$|\phi\left(r\right)|^{2}$ will be the potential seen by 
the axion string. Therefore, the axion string will prefer to move to a 
location where $|\phi\left(r\right)|^{2}$ is minimised, and that place 
is the core of the Higgs string. This is shown schematically in 
Fig.~\ref{bound_state_potential}. 
 
\begin{figure}[htbp] 
  \begin{center} 
    \includegraphics[width=0.8\textwidth,angle=0]{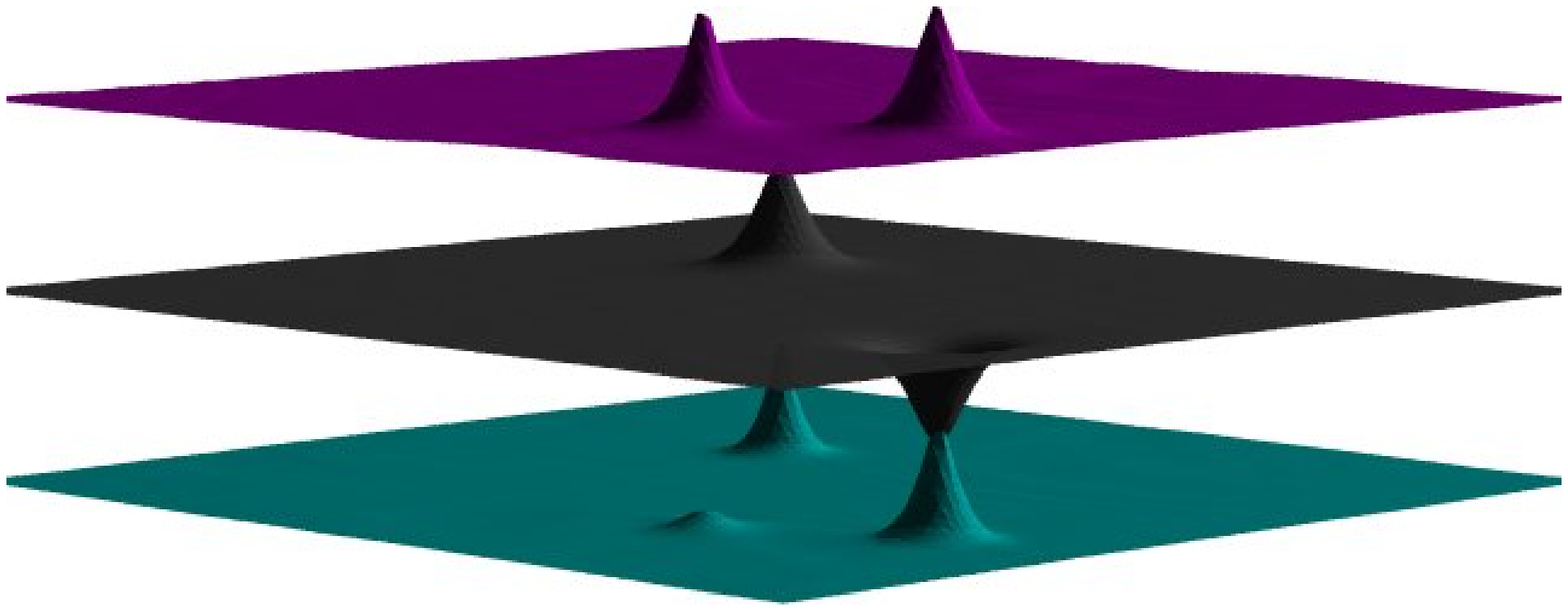} 

    \vspace {18pt} 

    \includegraphics[width=0.41\textwidth,angle=0]{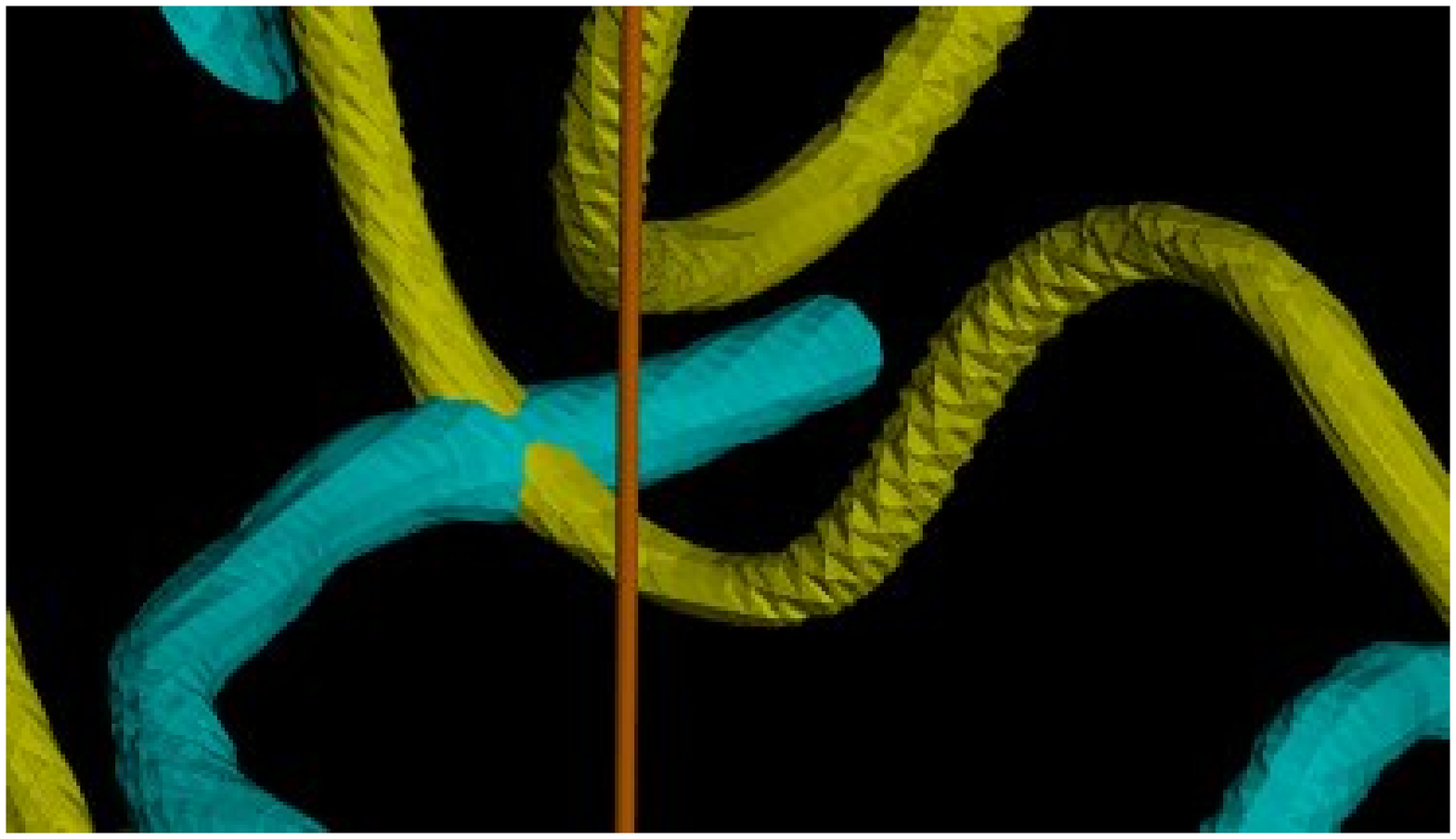} 
    \hspace {12pt} 
    \includegraphics[width=0.39\textwidth,angle=0]{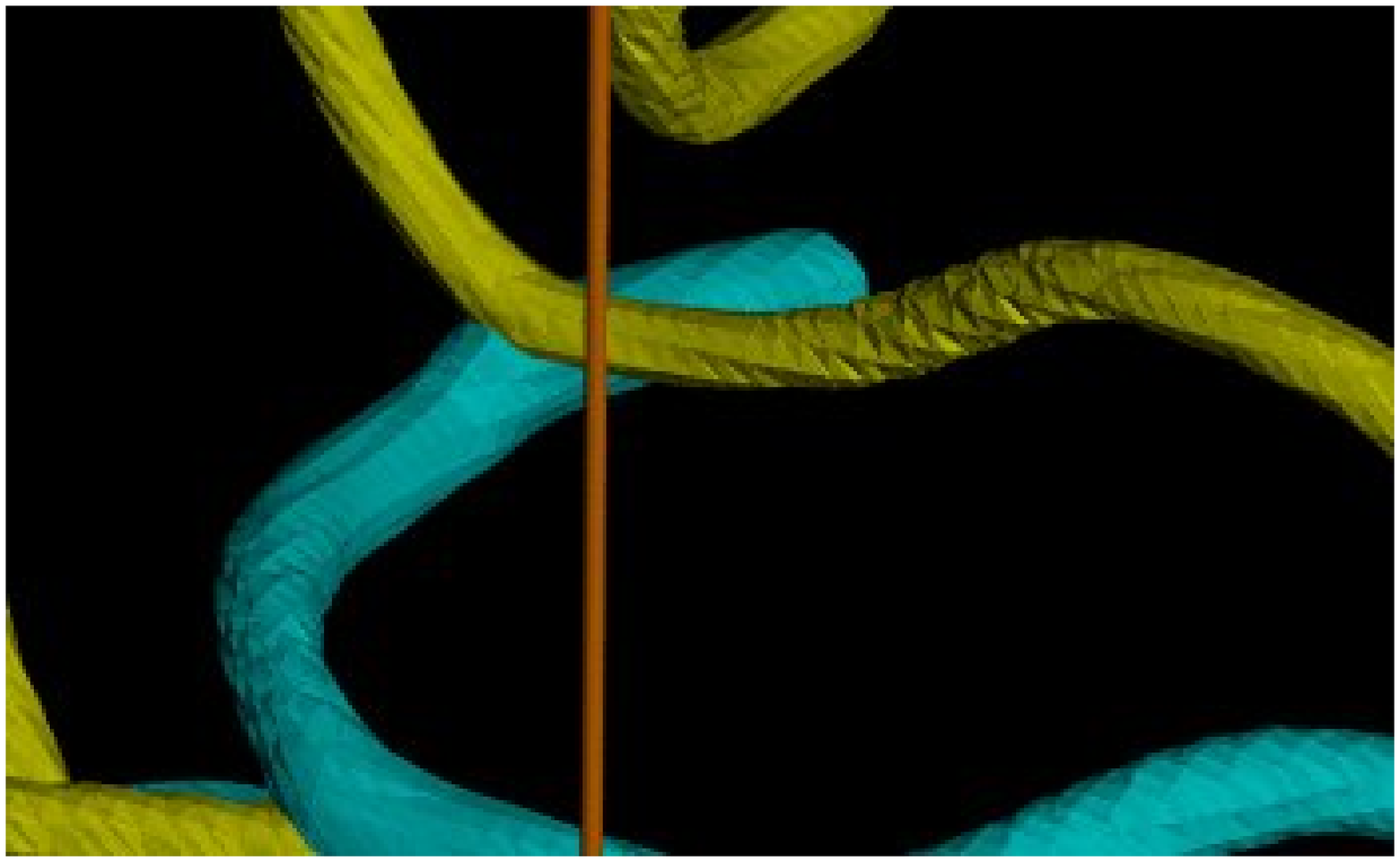} 
  \end{center} 
  \caption{Top: Bound state of vortices in two dimensions. From top to bottom, the layers show the axion field $|\chi$, the magnetic flux $\vec{\nabla}\times\vec{A}$ that couples to the Higgs, and the Higgs field $|\phi|$. The location of the defect on the right coincides in the Higgs and the axion fields, suggesting that they form a bound state. The binding is evident from the way the axion vortex drags the Higgs vortex with it, as it is pulled towards the anti-vortex on the left. The local Higgs vortex on the background does not move as its interactions with the other vortices are exponentially suppressed.  
  Bottom: In three dimensions, bound states split because of long-range interactions and the motion of the strings. As a result, bound states constitute only a small fraction of the entire string network. The two pictures show a bound state in a process of splitting.\label{2D_p_q_vortex}}
 \end{figure}

In an actual simulation we observe that when two strings intersect, 
they remain ``snagged'' and then the segments coming into the junction 
align themselves to allow the bound state segment to grow in length. 
The two processes working against this tendency are the long-range 
interactions between the strings and the motions of the unbound parts of the strings.
Snapshots of such bound string states, 
obtained in our numerical experiments, are shown in 
Fig.~\ref{2D_p_q_vortex}. 
 
\section{The $\left(p,q\right)$ string spectrum} 
 
The spectrum of the \pq strings was studied for various set-ups, some 
of which we would like to mention here. First of all, the most commonly 
used formula for \pq string tension, Eq.(\ref{pq_tension}), represents 
the BPS bound for an object carrying the charges of $p$ $F$-strings 
and $q$ $D$-strings, the more general result being that the mass per 
unit length of such an object satisfies the inequality (see, 
Ref.~\cite{Johnson:2000ch}): 
\begin{equation} 
\frac{M}{L} \ge \frac{\sqrt{p^{2}+q^{2}/g_{s}^{2}}} 
{2\pi\alpha^{\prime}}~. 
\end{equation} 
If we consider a more realistic model with explicit stabilisation of 
the compactification, the strings formed at the end of brane 
inflation will be localised at the bottom of a warped throat 
~\cite{Copeland:2003bj}, since the tension of the cosmic string is 
given by the value of the warp factor~\cite{Kachru:2003sx}: 
\begin{equation} 
\mu_{1} = \frac{1}{2\pi g_{s}\alpha^{\prime}}\left(\frac{r_{0}}{R} 
\right)^2~. 
\end{equation} 
It was noted in Ref.~\cite{Gubser:2004qj} that the D1 branes placed at  
the bottom of a warped throat embedded in a flux compactification  
cannot be BPS.  
For such an object with $q$ units of $D1$ charge the expression of the 
tension was found to be~\cite{Herzog:2001fq}: 
\begin{equation} 
\mu_{1, q} \sim \sin\left(\frac{\pi q}{M}\right)~,  
\end{equation} 
where $M$ is the number of units of R-R flux through the ${\bf S}^{3}$ 
at the bottom of a K-S throat, ~\cite{Klebanov:2000hb}. 
 
A different class of objects with $p$ $F$-strings and 
$q$ $D$-strings units of charge was studied in  
~\cite{Firouzjahi:2006vp}. Such an object is a $D3$-brane  
wrapped on the collapsing ${\bf S}^{2}$ of a K-S throat, and the  
spectrum of tensions for such an object is: 
\begin{equation} 
\mu_{\left(p, q\right)} = \frac{1}{2\pi\alpha^{\prime}} 
\sqrt{\left(\frac{bM}{\pi}\right)^{2} 
\sin\left(\frac{\pi\left(p-qC_{0}\right)}{M}\right)^{2} +  
\frac{q^{2}}{g_{s}^{2}}} 
\label{non_BPS_spectrum}~. 
\end{equation} 
Thus, the most commonly used formula for the spectrum of the \pq 
string tension, Eq.(\ref{pq_tension}), is exact only in the BPS limit, 
while for more general situations we expect departures from this 
theoretical prediction. In order to make a comparison of our model 
with the theoretical formula, Eq.~(\ref{pq_tension}), we 
will numerically determine the tension of the \pq strings for a range 
of values of the charges and then fit the numerical data with the 
expression given in Eq.~(\ref{pq_tension}).

The systems we use to model the \pq strings are different in a number 
of aspects from the ones used so far, and we believe that our systems 
bring a number of improvements and allow for additional effects to be 
studied. 
 
The systems used so far to simulate \pq string networks allow for the 
formation of 3-string~\cite{Hindmarsh:2006qn} (or more generally 
N-string~\cite{Copeland:2005cy}) junctions, in which all strings that 
meet at a point have equal tension.  In our models (constructed along 
the lines of ~\cite{Saffin:2005cs}) the bound states have 
different tension than the single-charge strings, following the 
original string theory model more closely.  Our models also allow to 
individually set the long-range interaction of each species of cosmic 
strings. As Eq.~(\ref{non_BPS_spectrum}) suggests, for cosmic strings 
at the bottom of a K-S throat the F-string is not BPS while the 
D-string is. We therefore expect the different components of the \pq 
string to exhibit different types of long-range interactions.  Also, 
as pointed out in Ref.~\cite{Copeland:2003bj}, in models where the 
standard model branes are located at orbifold fixed points, not all 
bulk fields that couple to the cosmic strings of the model survive the 
orbifold projection. There will also be other fields present, coming 
from higher-rank form-fields integrated over collapsed cycles, which 
will mediate the interactions of cosmic strings obtained from 
higher-dimensional branes wrapping the same collapsed cycles.  We 
therefore want to allow for the possibility to independently choose 
the long-range interactions between the different species of 
cosmic strings. 
 
\section{Numerical analysis of the $\left(p,q\right)$ string spectrum} 
 
In this section we construct the spectrum of $\left(p,q\right)$ 
strings formed in our model and compare it with the one predicted by 
string theory, Eq.(\ref{pq_tension}). We will start by considering an 
isolated defect and make an ansatz for the radial field profiles in 
analogy with the NO vortex~\cite{Nielsen:1973cs}. 
In fact, the model consists of 
two NO vortices coupled via the scalar field potential. 
 
The cosmic string we consider is an infinitely long, straight cosmic 
string, so we will choose cylindrical coordinates with the $z$ axis 
oriented along the string. The field configuration is independent of 
the coordinate along the string, $z$, so we are effectively describing 
a 2-dimensional vortex. 
 
The static, cylindrically symmetric ansatz for the fields of a string with 
charges $\left(p,q\right)$ is that of two NO vortices with winding numbers 
$p$ and $q$: 
\begin{eqnarray} 
\phi\left(r, \varphi\right)  &=&  
f\left(r\right)e^{-{\mathbbm i}p\varphi}\nonumber \\ 
\vec{A}\left(r\right) &=&  
\frac{a\left(r\right)}{r}\hat{e}_{\varphi}\nonumber \\ 
\chi\left(r, \varphi\right)  &=&  
h\left(r\right)e^{-{\mathbbm i}q\varphi}\nonumber \\ 
\vec{C}\left(r\right) &=& \frac{b\left(r\right)}{r}\hat{e}_{\varphi}~,  
\label{field_ansatz} 
\end{eqnarray} 
where $\vec{A}$ and $\vec{C}$ denote the spatial components of the
gauge fields.  The corresponding equations of motion are:
\begin{eqnarray} 
\frac{\partial^{2}a}{\partial r^{2}}- 
\frac{1}{r}\frac{\partial a}{\partial r}- 
e_{1}\left(p-e_{1}a\right)f^{2} &=& 0 \nonumber \\ 
\frac{\partial^{2}f}{\partial r^{2}}+ 
\frac{1}{r}\frac{\partial f}{\partial r} 
-\frac{\left(p-e_1a\right)^{2}}{r^2}- 
\lambda_{1}\left(f^{2}-\eta_{1}^{2}\right)f- 
\frac{\lambda_{2}}{2}\left(h^{2}-\eta_{2}^{2}\right)^{2}f 
&=& 0 \nonumber \\ 
\frac{\partial^{2}b}{\partial r^{2}}- 
\frac{1}{r}\frac{\partial b}{\partial r}- 
e_{2}\left(q-e_{2}b\right)h^{2} &=& 0 \nonumber \\ 
\frac{\partial^{2}h}{\partial r^{2}}+ 
\frac{1}{r}\frac{\partial h}{\partial r} 
-\frac{\left(q-e_{2}b\right)^{2}}{r^2}- 
\lambda_{2}\left(h^{2}-\eta_{2}^{2}\right)f^{2}h 
&=& 0 \label{eq_of_motion}~. 
\end{eqnarray}  
From the above equations, we see that at $r=0$ all four functions  
$a\left(r\right)$, $b\left(r\right)$, $f\left(r\right)$ and  
$h\left(r\right)$ vanish, while at $r=\infty$ the  
functions go asymptotically to: 
\begin{equation} 
a\left(r\right) \rightarrow \frac{p}{e_{1}} \;,\; 
f\left(r\right) \rightarrow \eta_{1} \;,\; 
b\left(r\right) \rightarrow \frac{q}{e_{2}} \;,\; 
h\left(r\right) \rightarrow \eta_{2}~.  
\label{asymptotics} 
\end{equation} 
We are interested in obtaining the energy per unit length  
of the bound states, so it will be more convenient to calculate 
the expression of the Hamiltonian in terms of the 4 functions given  
above, and perform a gradient flow to obtain the minimum energy  
configuration. The radial profiles of the fields will appear as a  
by-product of the energy minimisation process.  
 
\begin{figure}[htbp] 
  \begin{center} 
    \includegraphics[width=0.55\textwidth,angle=0]{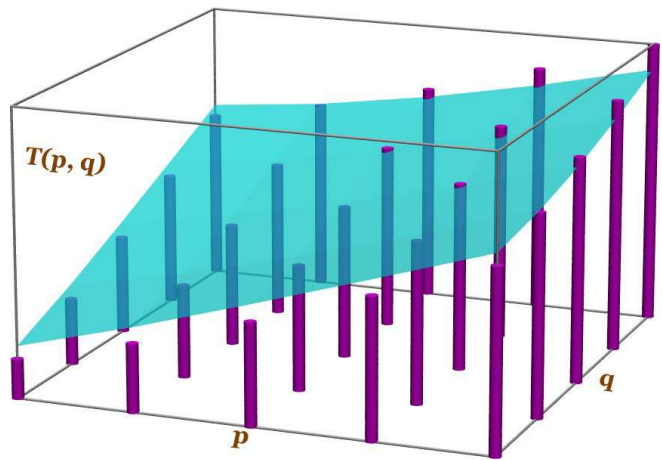} 
    \caption{Fit of the square-root spectrum. The vertical bars are the 
      string tension values calculated via the energy minimisation method. 
      The surface is the function $T\left(p,q\right)=\sqrt{ap^2+bq^2}$
      computed with the parameters 
      $a$ and $b$ determined from the fit. 
      \label{Fit_Spectrum}} 
  \end{center} 
\end{figure}

The Hamiltonian density has the expression: 
\begin{eqnarray} 
\mathcal{H}=\hf\left|\frac{\partial f}{\partial r}\right|^{2}+ 
\hf\left(p-e_{1}a\right)^{2}\frac{f^{2}}{r^{2}}+ 
\hf\left|\frac{1}{r}\frac{\partial a}{\partial r}\right|^{2}+ 
\frac{\lambda_{1}}{4}\left(f^{2}-\eta_{1}^{2}\right)^{2} 
\nonumber \\ 
 ~~~~+\hf\left|\frac{\partial h}{\partial r}\right|^{2}+ 
\hf\left(q-e_{2}b\right)^{2}\frac{h^{2}}{r^{2}}+ 
\hf\left|\frac{1}{r}\frac{\partial b}{\partial r}\right|^{2}+ 
\frac{\lambda_{1}}{4}f^{2}\left(h^{2}-\eta_{2}^{2}\right)^{2}~. 
\label{Hamiltonian} 
\end{eqnarray} 
In order to perform the gradient flow we start by discretising the 
radial direction into $N$ intervals. The 4 functions become 4 sets of 
$N+1$ variables whose values will need to be determined, so that the 
expression of the Hamiltonian integrated along the radial direction 
can be minimised. In the discrete version the integral is replaced by 
a sum, namely 
\begin{eqnarray} 
\fl H =2\pi\int_{0}^{\infty}r{\rm d}r\mathcal{H} 
\left(r\right)\rightarrow&&\nonumber\\ 
&& 2\pi \sum_{k=0}^{N-1}\left\{ 
\left(k+\hf\right)\frac{\left[f_{k+1}-f_{k}\right]^{2}}{2\dx} 
\right. 
+\frac{\left[a_{k+1}-a_{k}\right]^{2}}{\left(2k+1\right)\dx^3} 
\nonumber\\ 
&&\ +
\left(k+\hf\right)\frac{\left[h_{k+1}-h_{k}\right]^{2}}{2\dx}+ 
\frac{\left[b_{k+1}-b_{k}\right]^{2}}{\left(2k+1\right)\dx^3} 
\nonumber\\ 
&&+2\pi \sum_{k=1}^{N-1}\left\{ 
\frac{\left[p-e_{1}a_{k}\right]^{2}}{2k\dx}f_{k}^{2}+ 
\right. 
\left. 
\right.
\frac{\left[q-e_{2}b_{k}\right]^{2}}{2k\dx}h_{k}^{2} 
\nonumber\\
&&\ \left.+\frac{\lambda_{1}k\dx}{4}\left[f_{k}^{2}-\eta_{1}^{2}\right]^{2}+ 
\frac{\lambda_{2}k\dx}{4}f_{k}^{2}\left[h_{k}^{2}-\eta_{2}^{2}\right]^{2} 
\right\}~.  \label{discretized_hamiltonian}
\end{eqnarray}  
In a numerical calculation  the radial direction cannot be infinite, 
so we will have to choose it to be large as compared to the size of the 
defect core. The values of the functions at the ends of the intervals  
$k=0$ and $k=N$ will not evolve, they will be kept fixed at their  
core and asymptotic values, respectively. To find the equations of motion  
for the rest of the $4N-4$ variables we calculate the variation of the  
Hamiltonian with respect to each variable for $p=1\dots N-1$. 
One can then find the minimal energy configuration by evolving the equations  
\begin{equation} 
\frac{\partial f_{p}}{\partial \tau} = -\frac{\delta H}{\delta f_{p}}  
\,\,\, \dots \,\,\, 
\frac{\partial b_{p}}{\partial \tau} = -\frac{\delta H}{\delta b_{p}}~, 
\label{discretized_hamilton_eq}
\end{equation} 
starting with an appropriate initial configuration and keeping the  
values of the functions at the end of the interval fixed. The initial  
configuration we use is a linear radial profile starting  
with zero at $r=0$ and ending with the corresponding asymptotic value  
at $r = N\dx$. 
The results, together with a fit of the form \ref{pq_tension} 
are shown in Fig.~\ref{Fit_Spectrum}. 

\section{Results from numerical experiments} 
\label{sec:results}

To investigate the formation and evolution of bound states in a string network, we carried out simulations of the classical time evolution on lattices with two and three spatial dimensions. We carried out these
simulations in pairs, using the same initial conditions and parameters
of the Lagrangian in both cases. In the first case, the Higgs and the
axion have both local U(1) symmetry.  In the second case, the axion
field has a global U(1) symmetry, and therefore the axion strings
display long-range interactions.

The details of the lattice discretisation of these equation are presented in
\ref{app:lattice}. We  chose all parameters of the Lagrangian to have 
``natural'' values,
\begin{equation} \eta_1 = 1,\; \lambda_1 = 1,\; e_1 = 1,\; \eta_2 = 1,\; \lambda_2
= 1,\; e_2 = 1~.  \end{equation} 
The lattice spacing was $\delta x=0.45$, roughly half the
characteristic length scale set by the masses of the scalar and gauge
field in the broken phase. The time step what $\delta t=10^{-2}$.
The simulations were carried out with three different box sizes $200^3$, $256^3$ and $400^3$.

We used initial conditions in which the gauge potentials $A_\mu$ and $C_\mu$
are zero everywhere, so we have 
vanishing electric and magnetic fields. We also set the initial values of 
the time derivatives of the scalar fields to zero. The values of the 
fields themselves were set at the maximum of the potential $\phi = 0$, 
$\chi = 0$. To this configuration we added, depending on the problem 
studied, a small-amplitude random white-noise component, or a 
small-amplitude component with phases chosen such that the 
network of cosmic strings will form in a particular configuration. 
For such an initial configuration the 
Gauss constraint~(\ref{equ:Gauss}) is automatically satisfied.  

The white-noise initial conditions correspond to the symmetric phase
with zero correlation length. As the system starts to evolve, the U(1)
symmetries get spontaneously broken, and strings are formed by the
Kibble mechanism~\cite{Kibble:1976sj}. Because the gauge fields were
set to zero, no flux trapping~\cite{Hindmarsh:2000kd} takes place, in
contrast with thermal initial conditions.  The initial white-noise
configuration excites all modes, with half-wavelengths up to the
lattice spacing, and in order to see the formation and evolution of a
network of defects we have to drain the excess energy from the system.
We achieved that by including a constant damping $\sigma=0.2$ 
term for the scalar
fields and an ohmic term for the gauge fields, such that both gauge
fields satisfy the Gauss constraint~(\ref{equ:Gauss}) during the entire simulation.

What we compare in the two networks is the fraction of the volume of
the simulation box occupied by bound states. For completeness, we also
track the fraction of the simulation box volume occupied by single
charged strings.  Our set-up allows us to start with initial conditions
which will lead to the formation of a particular network
configuration. Starting with a white-noise configuration, the total
volume occupied by bound states is too small to lead to a sizable
difference in their volumes, for the two types of string networks.
This is also due to the particular form of the potential for the axion
field. This potential vanishes at the core of a Higgs string and gets
its largest value where the Higgs field is in the vacuum. Therefore,
starting with white-noise initial conditions the axion strings prefer
to form where the Higgs field is already in the vacuum. Thus, the
formation of bound states is not favoured. Even when we use a very
large simulation box, $400^3$, we encountered situations during the
network evolution where no bound states were present.  It is therefore
essential to be able to choose the initial phases of the scalar fields
such that the phases have various degrees of ``alignment'', so that
the formation of bound states is favoured. In this sense, white-noise
corresponds to no alignment at all, while choosing identical initial
conditions for both the Higgs and the axion fields results in a
network with all strings formed in bound states.

Our aim was to observe the evolution of the network of \pq strings,
paying special attention to the evolution of bound states.  We
observed that in the case of local-global networks the bound states do
not have a significant influence on the evolution of the network. The
long-range interaction of the global strings prevents bound states of
significant size from being long lived.  This is not the case for
local-local string networks in which we observed long lived bound
state string segments. These bound states have a significant effect on
the evolution of the network.
 
\begin{figure}[htbp] 
  \begin{center} 
    \includegraphics[width=0.55\textwidth,angle=0]
{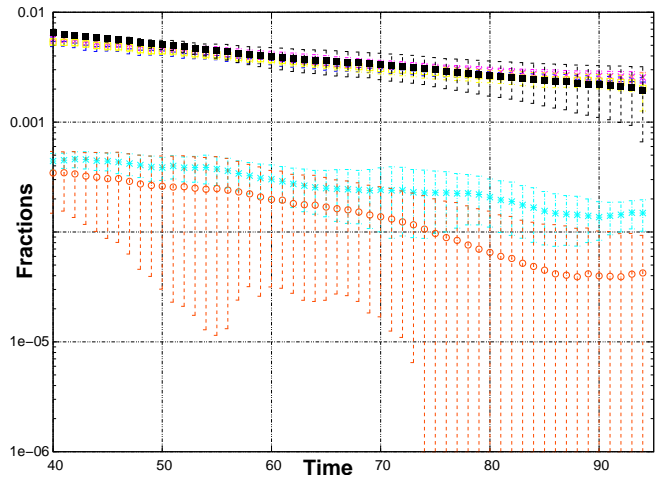}
    \caption{(color online) The volume of the bound states for local-local (blue) and
      local-global (orange) networks. The bound states of the
      local-global network show a much larger spread in volume as a
      consequence of their short lifetime. The curves at the top
      represent the volumes of the simulation box occupied by each
      species of cosmic string. In this example all simulations were
      performed in a $200^3$ box.  The volumes occupied by each type
      of string and the bound states are averaged over 10 different
      initial conditions.
      \label{horror_bars}}
  \end{center} 
\end{figure}

More quantitatively, Fig.~\ref{horror_bars} shows the fraction of the
simulation box occupied by the Higgs strings, the axion strings, and
their bound states. If we choose white-noise initial conditions the
bound states formed are longer-lived in the local-local network than
in the local-global one, but the small overall volume of the bound
states did not allow us to obtain results of any statistical
significance.
Nevertheless, we observed the same trend at all lattice sizes: 
The volume
occupied by the bound states is larger in the case of a local-local
network and the total energy is larger for the local-global network.
For the largest box we also encountered a situation in which the
network did not feature any bound states. While this was certainly a
finite-size effect, in the local-local network evolving from identical
initial conditions the bound states survive past the point where the
local-global bound states disappear.
 
However, if we choose the Higgs and the axion fields to be partially
aligned in the initial configuration the volume of the bound states
formed is large enough to obtain a statistically significant
difference. We obtain the partial alignment by choosing the initial
configurations identical and then rotating the one for the axion
field by $90^{o}$. The fields are identical along the rotation
axis. We performed a number of simulations starting from various
initial conditions and the results are summarised in
Fig.~\ref{horror_bars}.  We observe that the local-local network
forms bound states that are much longer-lived than those of the
local-global network. The short lifetime of the bound states of the
local-global network results in a much larger spread of the bound
state volume, and it is common to encounter situations where no bound
states are present.

\begin{figure}[htbp] 
\begin{center} 
\includegraphics[width=0.4\textwidth,angle=0]{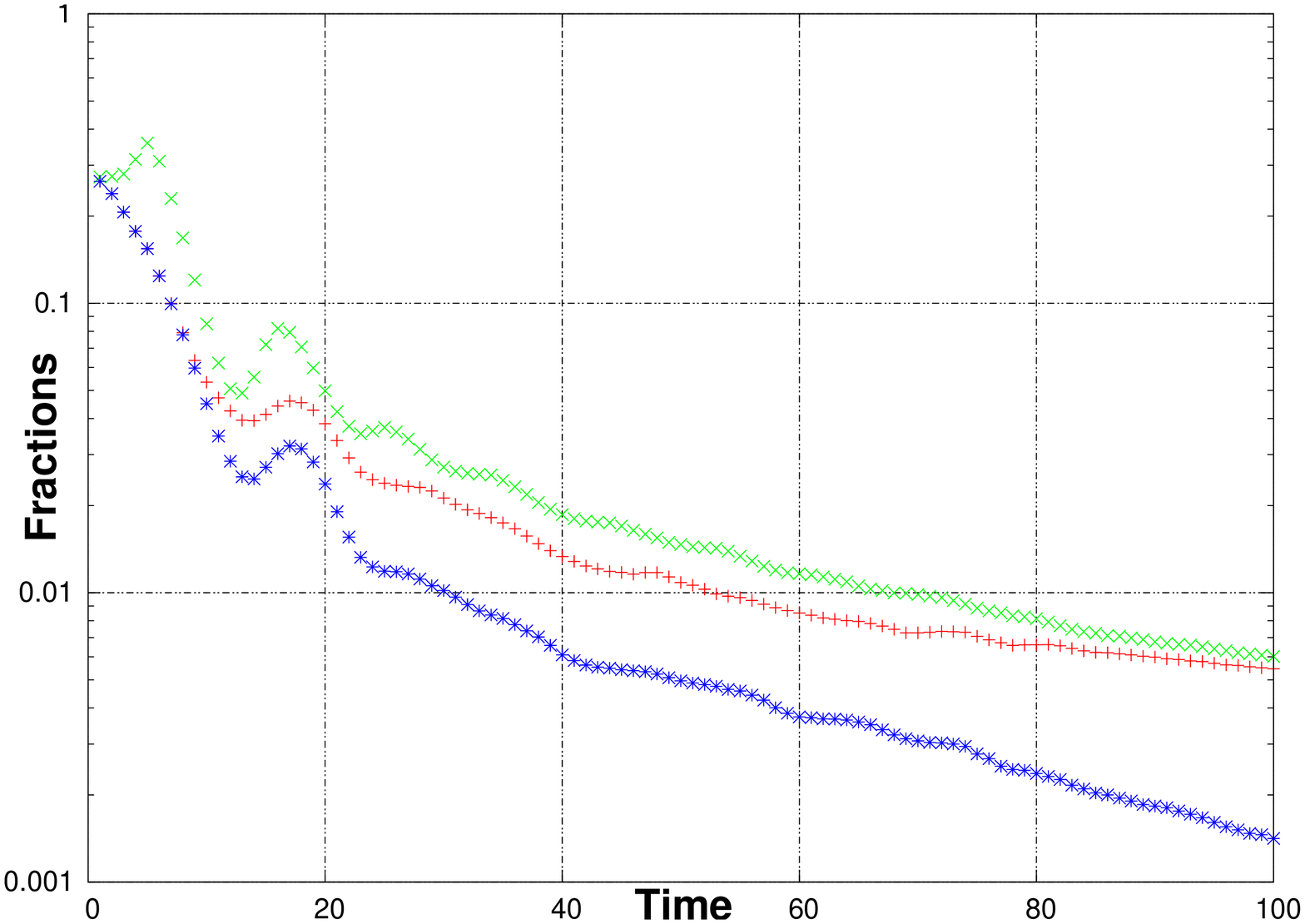}
\includegraphics[width=0.4\textwidth,angle=0]{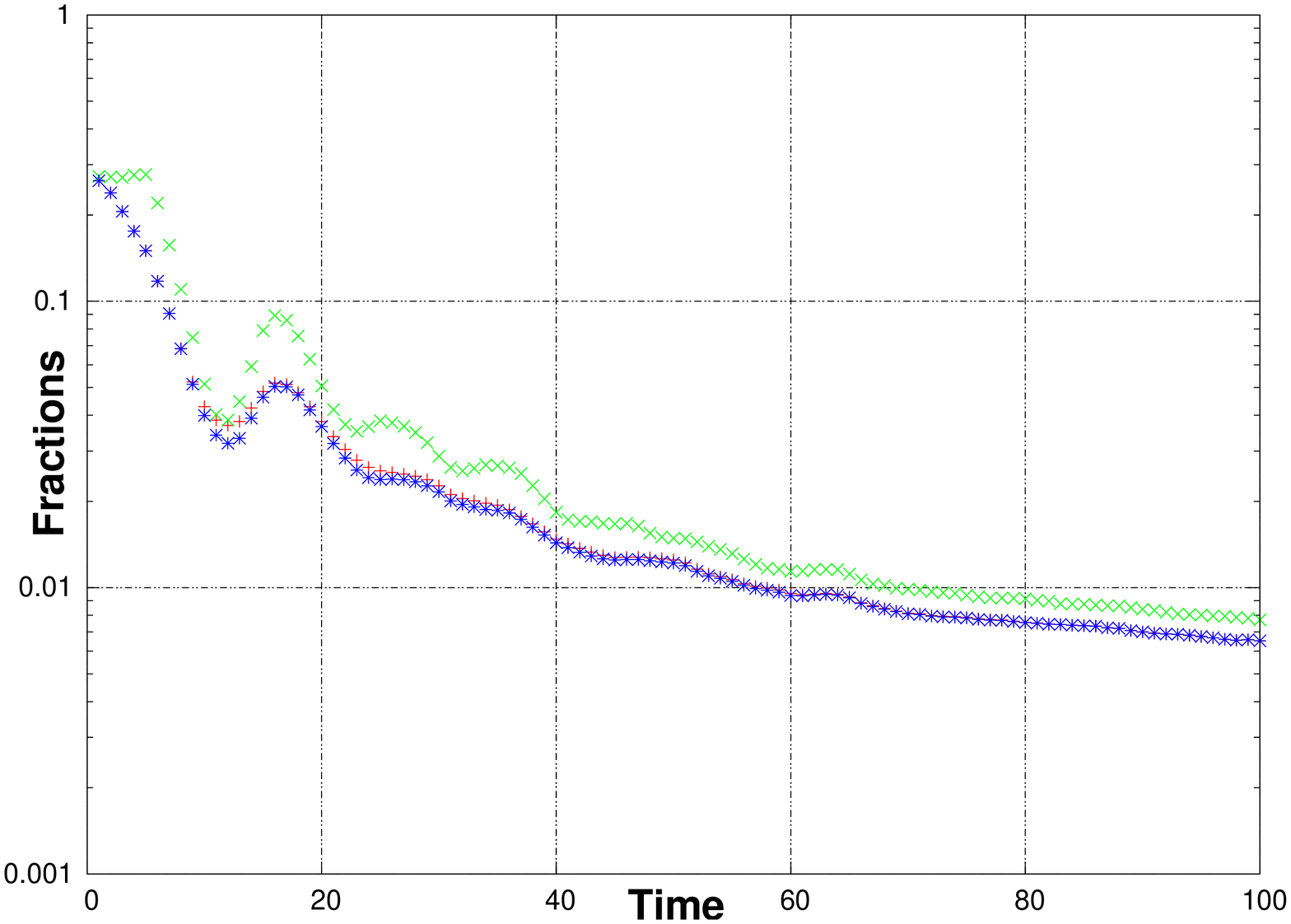}
\end{center} 
 
\begin{center} 
\includegraphics[width=0.4\textwidth,angle=0]{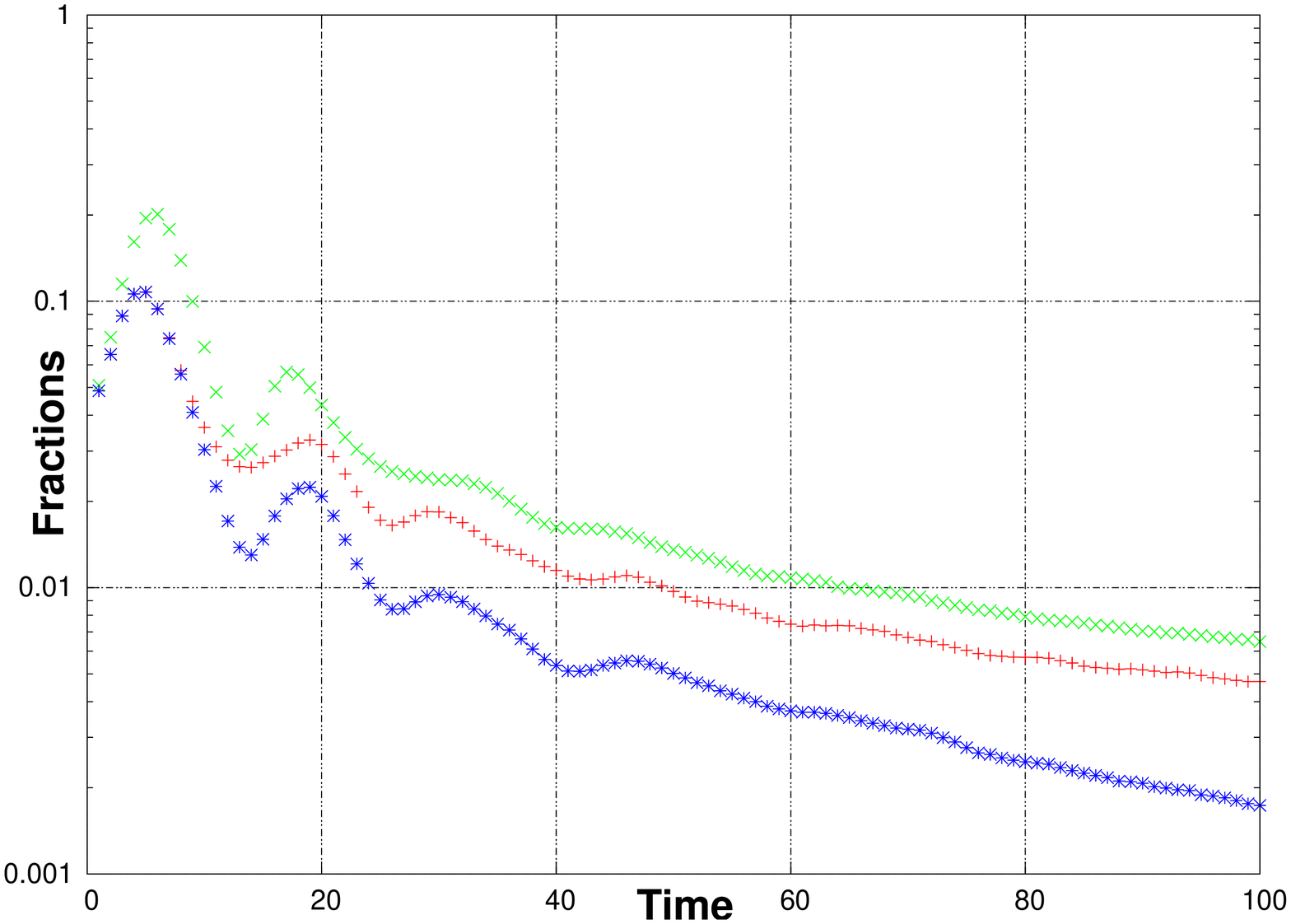}
\includegraphics[width=0.4\textwidth,angle=0]{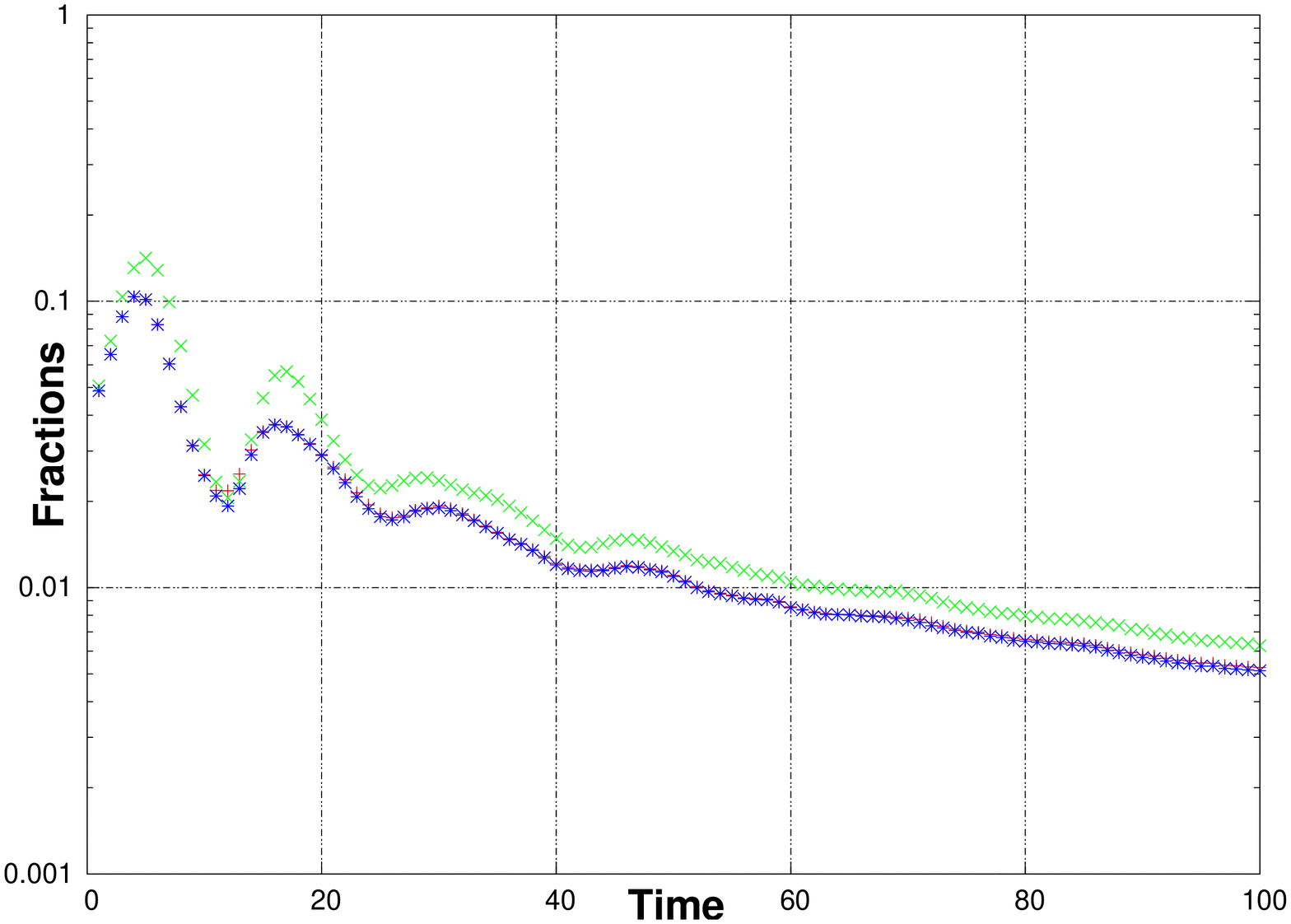}
\end{center} 
\caption{(color online) The total physical volume of the simulation box occupied by
  Higgs strings (green), the axion strings (red), and their bound
  states (blue).  On the left we show two examples of volume
  evolutions for local-global networks and on the right the
  corresponding values for a local-local network starting from the
  same initial conditions as the local-global ones.  We see that in the
  local-global case a significant fraction of the strings become
  unbound early in the simulation.  In the local-local case the volume
  of the bound states is identical with the one of the thinner (axion)
  strings.
  \label{reverse_problem}} 
\end{figure}

The most convincing evidence comes from analysing the reverse problem,
namely that of a bound state splitting as a result of the long-range
interactions between strings. We performed a number of simulations in
which we started with an already-formed network with all the strings
carrying $\left(1,1\right)$ charge. We achieved that by choosing
identical (``fully aligned'') initial configurations for the Higgs and
the axion fields. These results, presented in
Fig.~\ref{reverse_problem}, show most clearly that the effect of the
long-range interactions is to cause the bound states to split. In the
absence of long-range interactions the strings remain in the
$\left(1,1\right)$ state throughout their entire evolution. The total
physical volume occupied by the bound states is identical with the
volume of the thinner strings.  Due to our particular choice of
binding potential the strings do not have identical thickness. There
are only minor deviations in the initial phase of the evolution when
the strings are not yet fully formed.

The main result of the simulations is that the long-range interactions
between global strings make the bound states of the local-global
network short-lived. The bound states turn out not to have any
significant contribution to the evolution of the network; the
evolution is dominated by the long-range interactions between
strings. In the case of a local-local network, the absence of
long-range interactions allows for the formation of bound states of
larger volume and these bound states do have an important influence on
the evolution of the network.
 
\begin{figure}[htbp] 
  \begin{center} 
    \includegraphics[width=0.35\textwidth,angle=0] 
{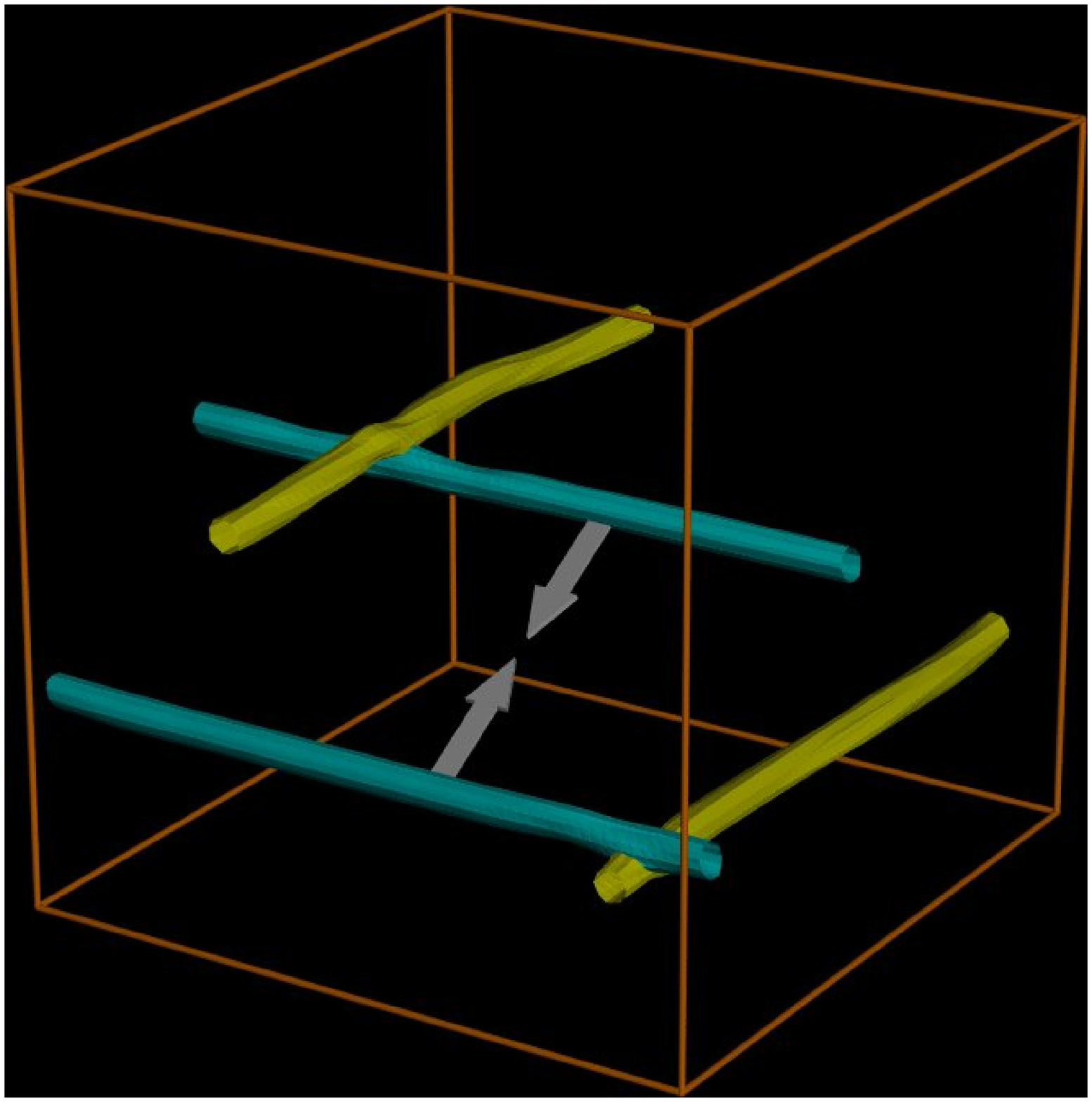} 
    \includegraphics[width=0.35\textwidth,angle=0] 
{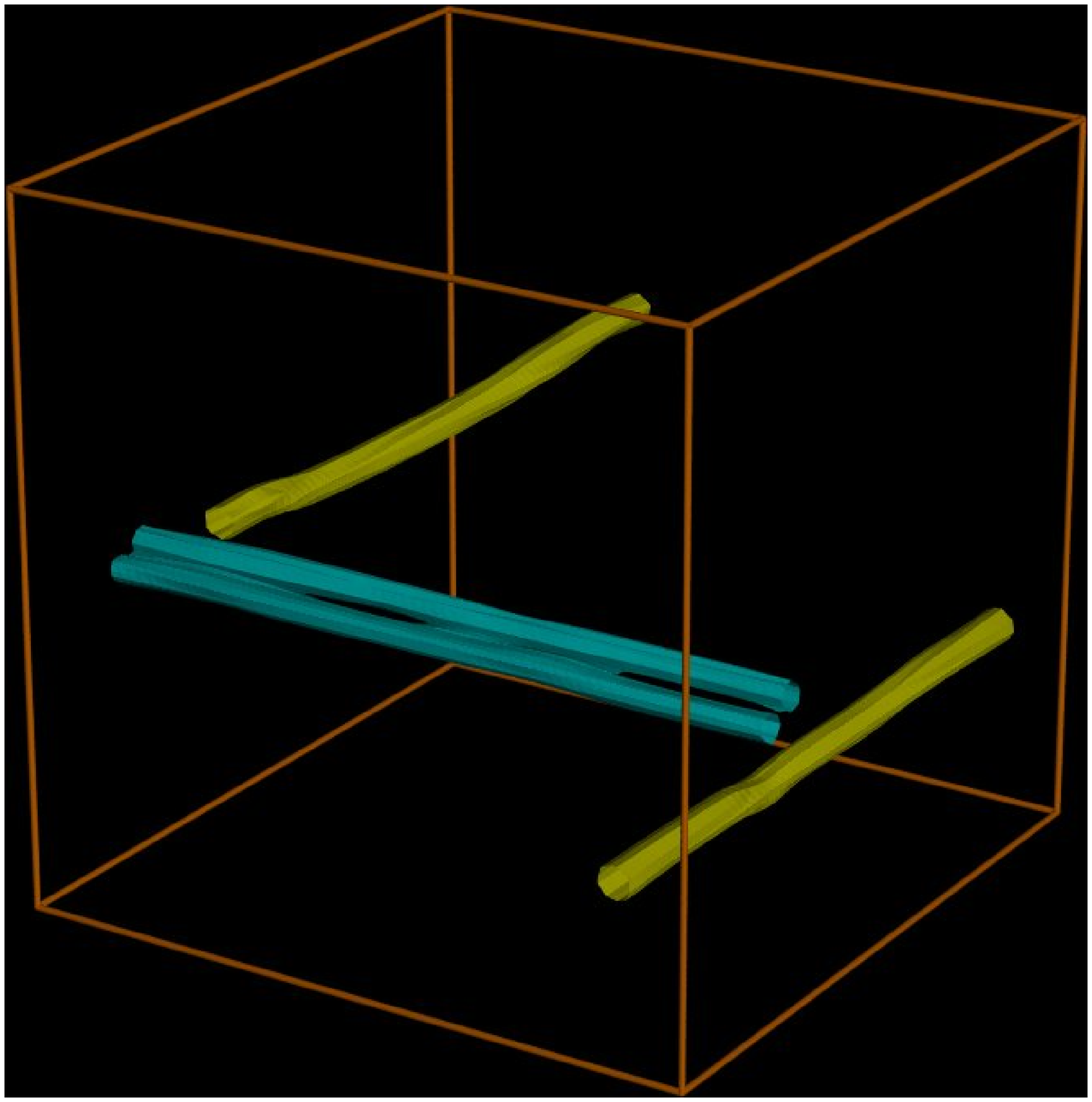} 
  \end{center} 
  \caption{ The bound states do not survive the long-range interaction  
   of the global strings.
    \label{split_bound_state}} 
\end{figure}

Another way to visualise the competing effects of the bound states and
the long-range interactions is presented in
Fig.~\ref{split_bound_state}. In this example we set the initial
conditions such that there was only one pair of local and one pair of
global strings. The strings form in pairs of opposite orientation
because of the periodic boundary conditions. We arranged the
configuration such that the attractive interactions between global
strings resulted in their motion towards the local ones, with the aim
to find out whether the formation of the bound states can stop the
motion of the global strings.
 
We performed the simulations in boxes of various sizes and with 
different winding numbers for the local strings. In each case we 
observed that the long-range interactions caused the bound states to 
split. The global strings moved towards the local ones and crossed them, 
forming bound states in this process. These bound states then split as 
the global strings continued to move towards each other. Finally they collided 
and annihilated.

\section{Conclusions} 
 
In  this paper we have analysed the evolution of \pq string 
networks in an attempt to 
observe the influence that the bound states have on the evolution of 
the network.  The \pq string network was modelled using two sets of 
Abelian Higgs fields. We studied two models, one in which both species 
of string have only short-range interactions and another one in which 
one species of string features long-range interactions.  Our aim was to 
observe the competition between the binding and the long-range 
interactions in an attempt to see which effect will have a bigger 
influence on the evolution of the string network. 
 
We started out with a simple set-up of local and global vortices in two 
space dimensions, where we could see that the two types of vortices 
form bound states and the long-range interactions between the global 
strings will drag the local vortices along with the global ones. The 
local vortices did not move as they displayed only short-range 
interactions. 
 
We then continued with a three-dimensional model in which we studied the  
evolution of cosmic superstrings starting from two types of initial 
conditions. In one case we looked at only two pairs of cosmic superstrings,  
one local and one global, oriented at right angles.  The initial positions  
were chosen such that the attractive interaction of the global strings 
would cause them to intersect the local ones. This way we could see whether 
the long range interaction will overcome the formation of bound states of the 
two string species. We repeated the simulation choosing local strings with 
different windings around the periodic simulation box, such that the local and 
the global strings would intersect in more than one point. In both cases we 
observed that the long-range interaction wins in the sense that the bound 
states split and the global strings move towards each other and annihilate.  
 
The other case we studied was the formation and evolution of a network  
starting with various initial conditions. The evolution of 
the string networks suggests that the long-range interactions have a much 
more important r\^ole in the network evolution than the formation of bound 
states. In the  local-global networks the bound states tend to split  
as a result of the long-range interactions, resulting in two networks 
that evolve almost independently. The formation of short-lived bound states 
and their subsequent splitting only increases the small-scale ``wiggliness'' 
of the local strings. In the case of a local-local network, the absence of  
long-range interactions allows the bound states to be much longer-lived and  
significantly influences the evolution of the string network.  
 
\begin{figure}[htbp] 
\begin{center} 
\includegraphics[width=0.36\textwidth,angle=0]{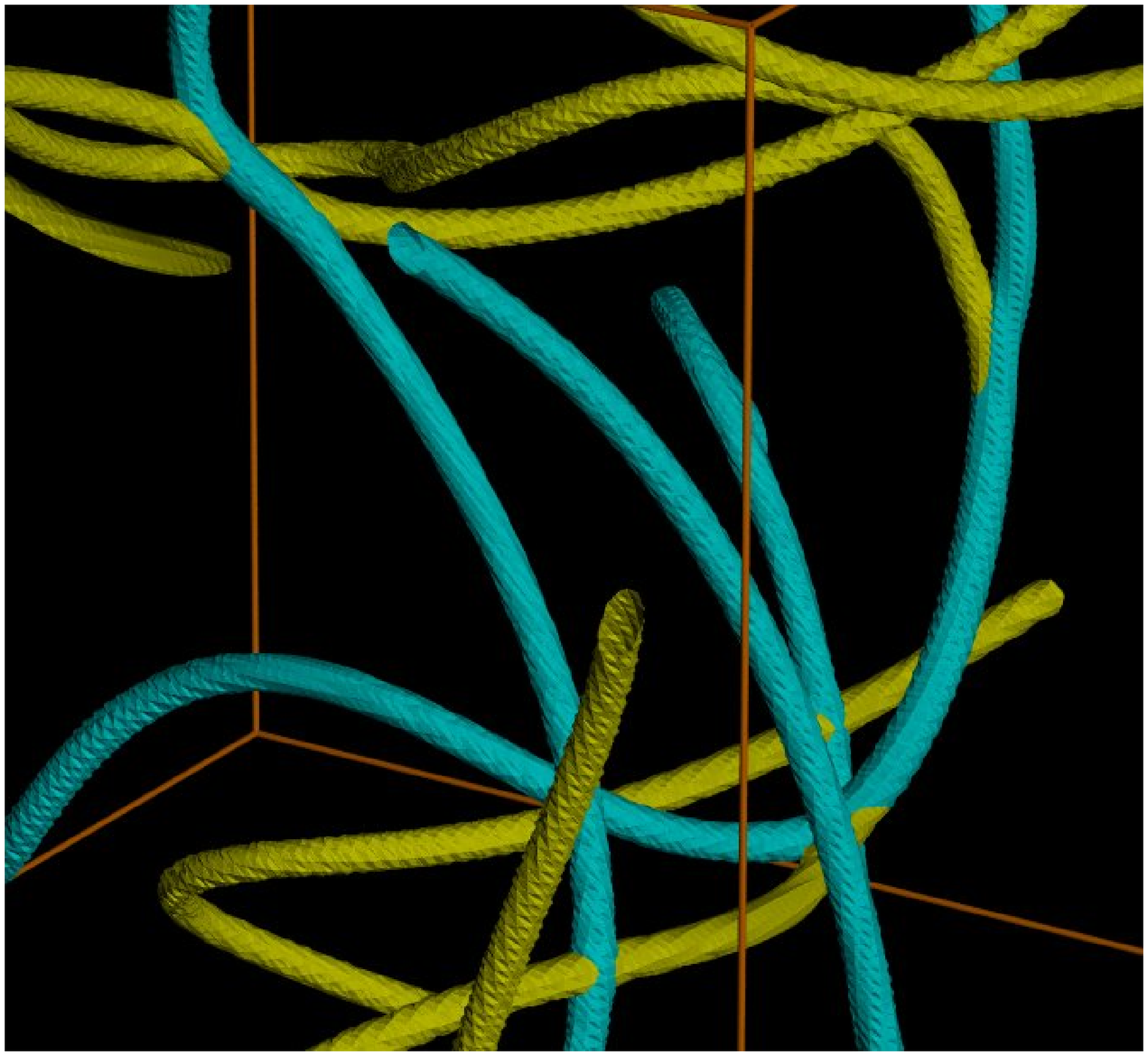} 
\includegraphics[width=0.4\textwidth,angle=0] {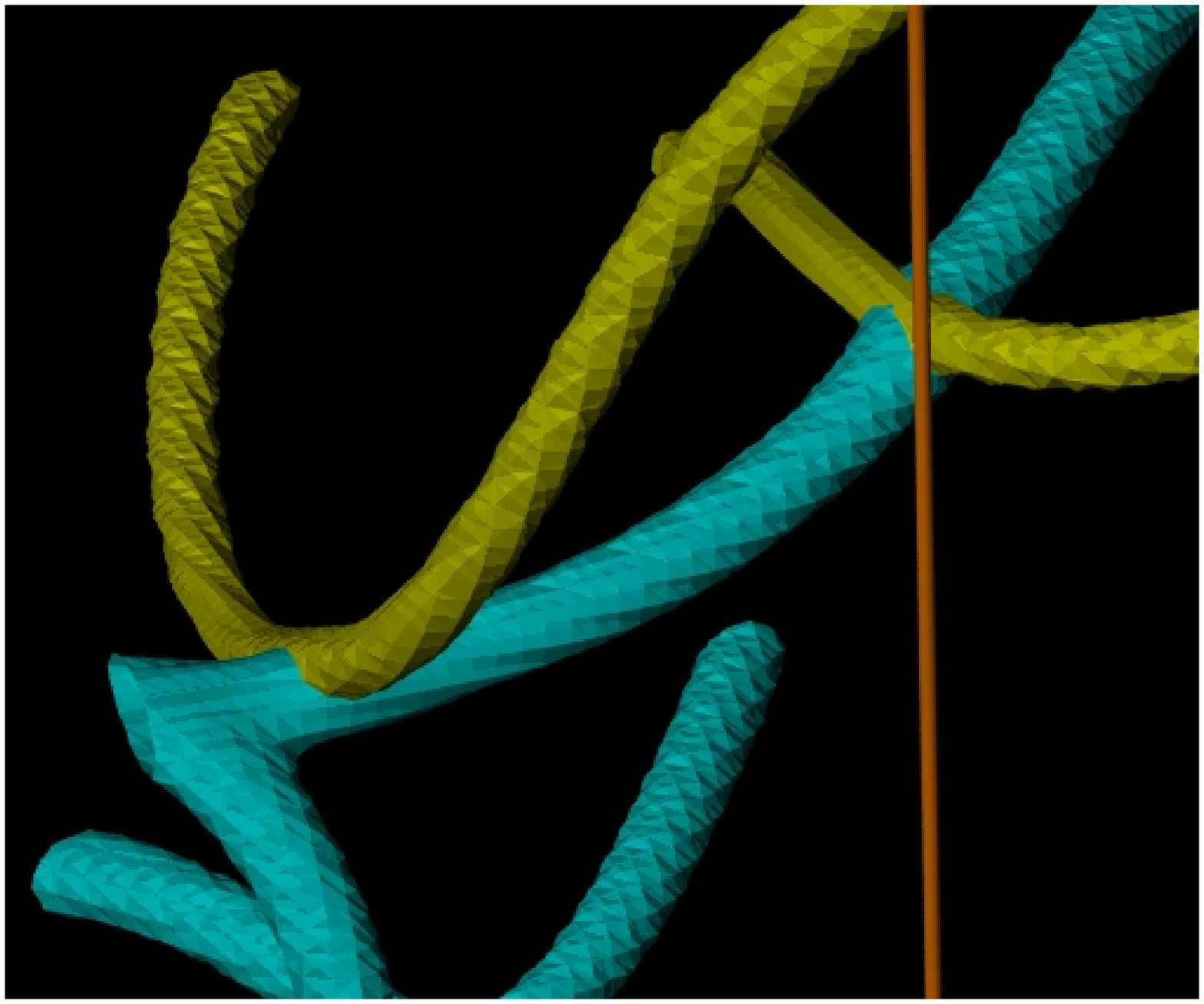} 
\end{center} 
\caption{Left: Bound states for local-local $\left(p,q\right)$ 
  strings. Right: Bound states for local-global $\left(p,q\right)$  
  strings. 
  \label{bound_states_pictures}} 
\end{figure} 

In an attempt to express these observations in a more quantitative
manner we plotted the fraction of the spatial volume occupied by the
bound states for each type of a network, as well as the total energy
of the string network as a function of time.  The observations we
mentioned above translate into a larger volume occupied by the bound
states, in the absence of long-range interactions. The energy of the
local-global network is larger than that of the local-local one, since
the energy of the global strings is not localised and therefore the
formation of bound states has a smaller effect on the total energy of
the network.
 
Our results suggest that if in a \pq string network one species of
cosmic string features long-range interactions, the bound states will
not survive and we have two networks evolving (almost)
independently. We did not address the question of scaling in this
work, but since the bound states split due to the long-range
interactions, we expect such a network to scale rather than freeze.
 
Our study included only two extreme cases, that of local-local and of
local-global networks. While we believe that the models we use
represent an improvement over the previous types of models considered,
as far as the tension of the strings and their bound states is
concerned, there is still room for improvement in modelling the
long-range interactions.
 
As it was shown in Ref.~\cite{Moore:2006ec}, the long-range
interactions between the cosmic strings can be much more accurately
modelled by including a rank-2 antisymmetric tensor field that couples
to the magnetic flux trapped at the core of the string via a
Chern-Simons term. It was also shown that, depending on the strength
of the Chern-Simons coupling, the interaction mediated by the rank-2
tensor field interpolates between that of local and global
defects. When the coupling is very small the interaction is
negligible, while at strong coupling the magnetic flux at the core of
the string is almost eliminated and the network behaves like a global
string network. We therefore expect the interplay between the effects
of binding and those of long-range interactions to be more subtle as
one changes the strength of the Chern-Simons coupling. At small
coupling we expect the evolution of the network to be dominated by the
formation of bound states, while at large coupling the effect of bound
states is negligible, and one gets two species of strings evolving
almost independently.
 
We believe that the models presented in this paper represent a step
forward in understanding the behaviour of a cosmic superstring network
formed at the end of brane inflation in flux compactification
scenarios. In a future work, we plan to use the models presented above
to address the issue of scaling in a \pq string network.

\ack 

It is a pleasure to thank E.\ Copeland, R.\ Gregory, A.\ Padilla and  
P.\ Saffin for discussions. The work of A.\ R. and H.\ S. was supported 
in part by the EU under MRTN contract MRTN-CT-2004-005104
and by PPARC under rolling grant PP/D0744X/1. 
The research of M.\ S. was supported in part by the European Union
through the Marie Curie Research and Training Network UniverseNet
(MRTN-CN-2006-035863).
 
\appendix
\section{Equations for energy minimization}

Varying the Hamiltonian Eq.(\ref{discretized_hamiltonian}) 
with respect to the values of the fields
at each point along the discretized radial direction one 
obtains:

\begin{eqnarray} 
&& \frac{\delta H}{\delta f_{p}} =   
\left(p-\hf\right)\frac{f_{p}-f_{p-1}}{\dx}+ 
\left(p+\hf\right)\frac{f_{p}-f_{p+1}}{\dx} 
\nonumber \\ 
&&~~~~~~+ \frac{\left[n-e_{1}a_{p}\right]^{2}}{p\dx}f_{p}+ 
 p\dx\lambda_{1}\left[f_{p}^{2}-\eta_{1}^{2}\right]f_{p}+ 
\frac{p\dx\lambda_{2}}{2}\left[h_{p}^{2}-\eta_{2}^{2}\right]^{2}f_{p}~,\\ 
&& \frac{\delta H}{\delta a_{p}} = 
-\frac{e_{1}\left[n-e_{1}a_{p}\right]}{p\dx}f_{p}^{2} 
+\frac{1}{p\dx^{3}}\left[\frac{a_{p}-a_{p-1}}{p-\hf}+ 
\frac{a_{p}-a_{p+1}}{p+\hf}\right]~, 
\\ 
&& \frac{\delta H}{\delta h_{p}} =   
\left(p-\hf\right)\frac{h_{p}-h_{p-1}}{\dx}+ 
\left(p+\hf\right)\frac{h_{p}-h_{p+1}}{\dx} 
\nonumber \\ 
&&~~~~~~+ \frac{\left[m-e_{2}b_{p}\right]^{2}}{p\dx}h_{p}+ 
p\dx\lambda_{2}\left[h_{p}^{2}-\eta_{2}^{2}\right]f_{p}^{2}h_{p}~, 
\\ 
&&  \frac{\delta H}{\delta b_{p}} = 
-\frac{e_{2}\left[m-e_{2}b_{p}\right]}{p\dx}h_{p}^{2} 
+\frac{1}{p\dx^{3}}\left[\frac{b_{p}-b_{p-1}}{p-\hf}+ 
\frac{b_{p}-b_{p+1}}{p+\hf}\right]~. 
\end{eqnarray}  

\begin{figure}[htbp] 
\begin{center} 
\includegraphics[width=0.45\textwidth,angle=0]{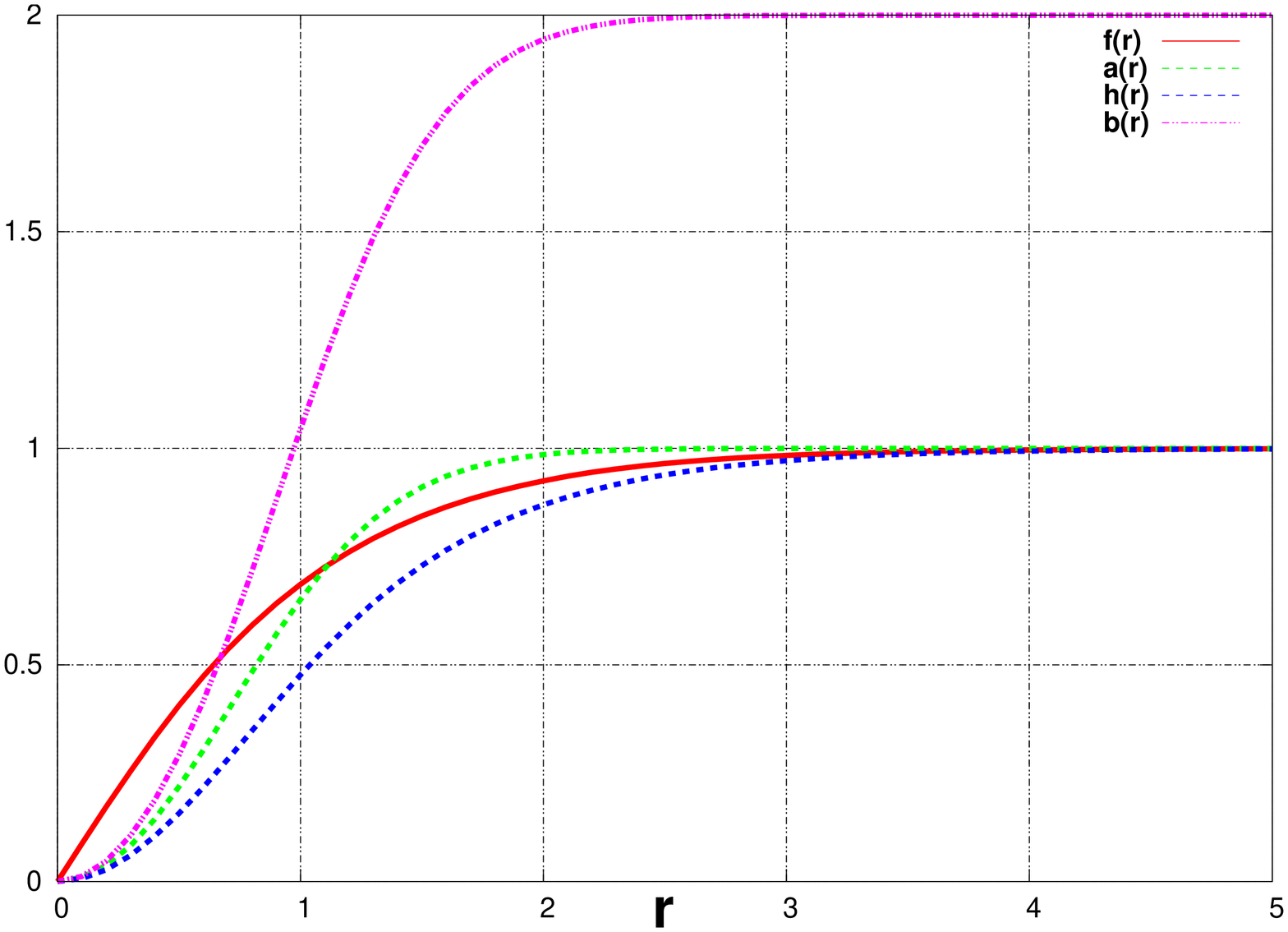} 
\includegraphics[width=0.45\textwidth,angle=0]{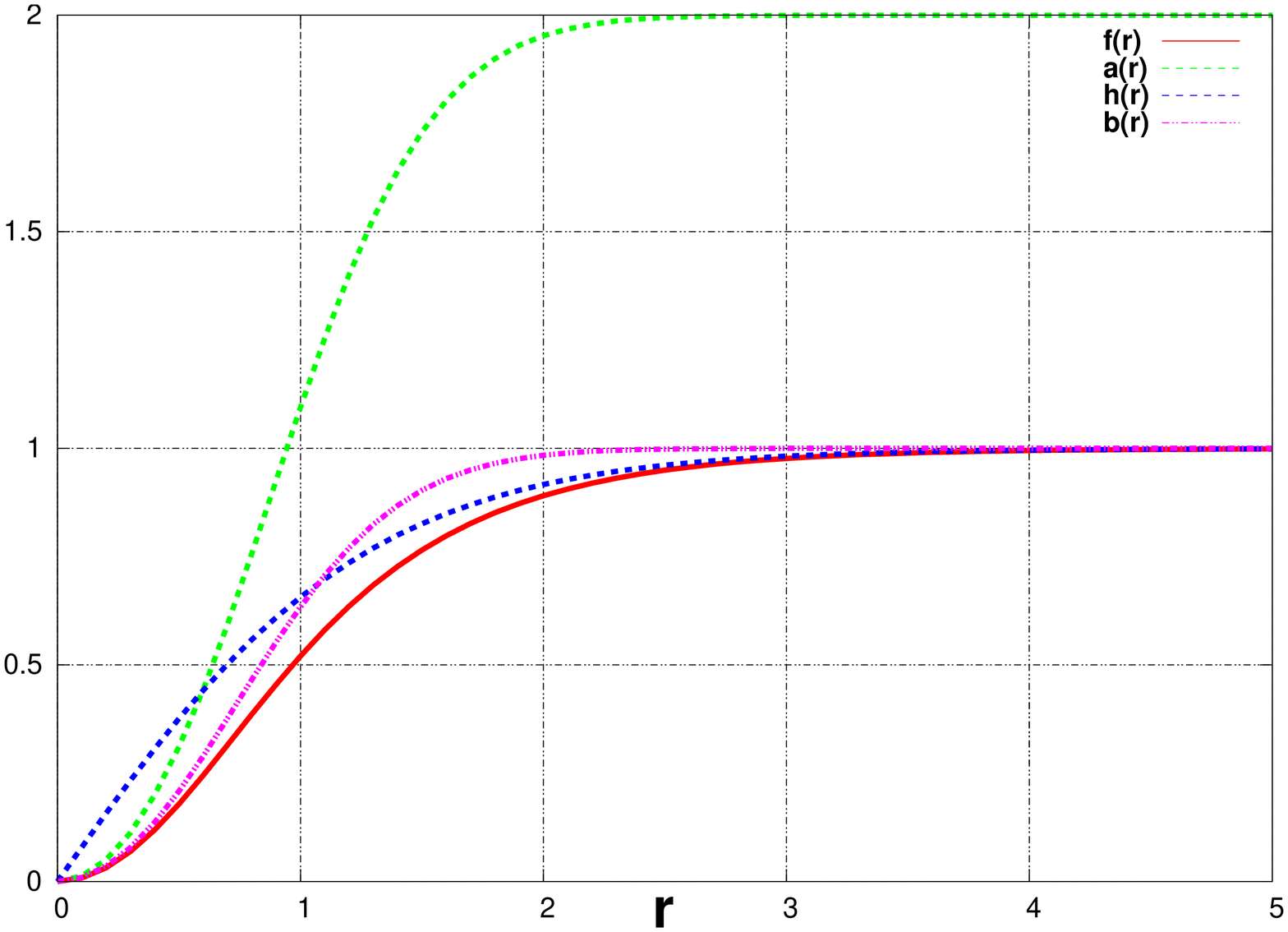} 
\end{center} 
 
\begin{center} 
\includegraphics[width=0.45\textwidth,angle=0]{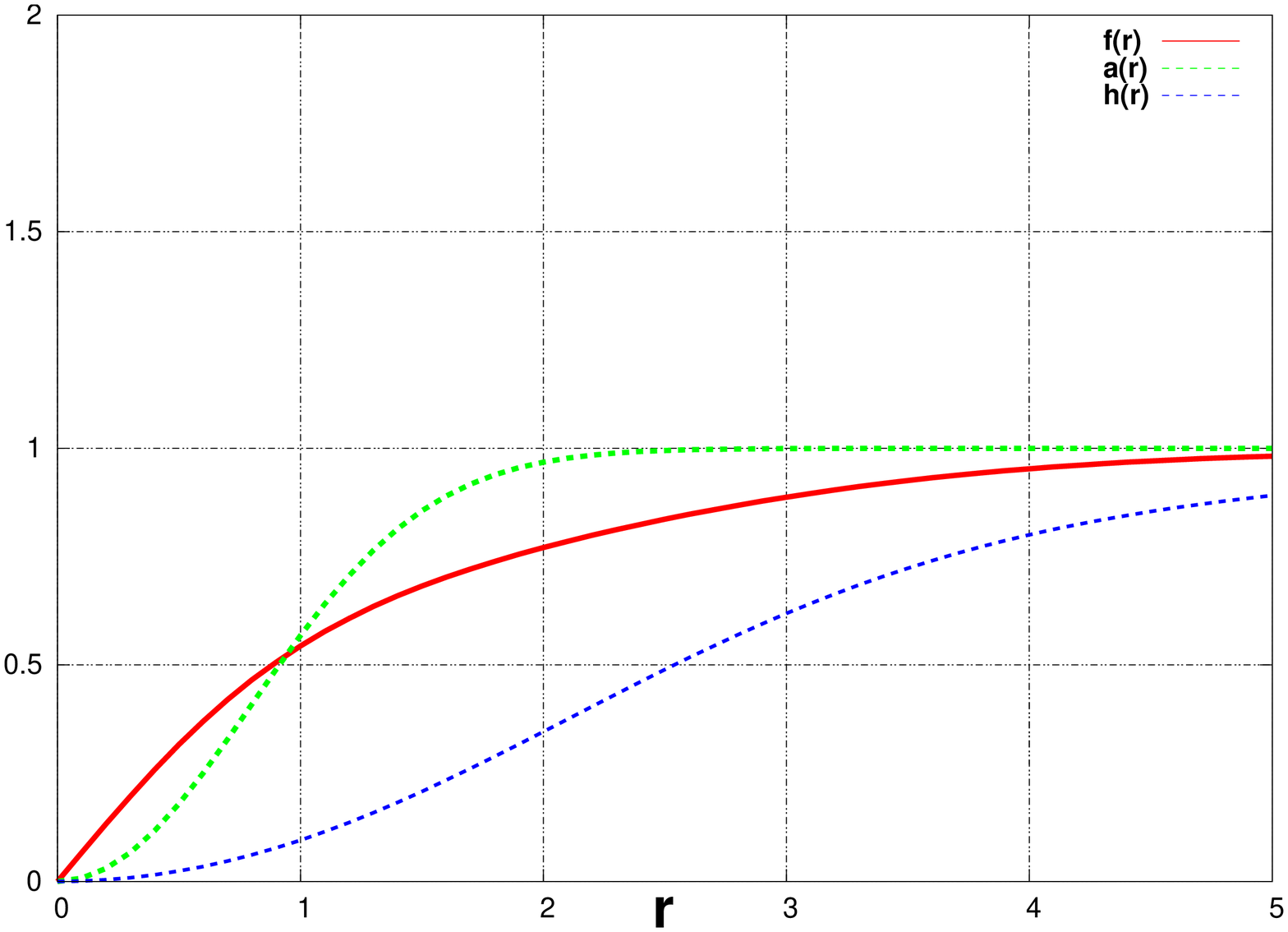} 
\includegraphics[width=0.45\textwidth,angle=0]{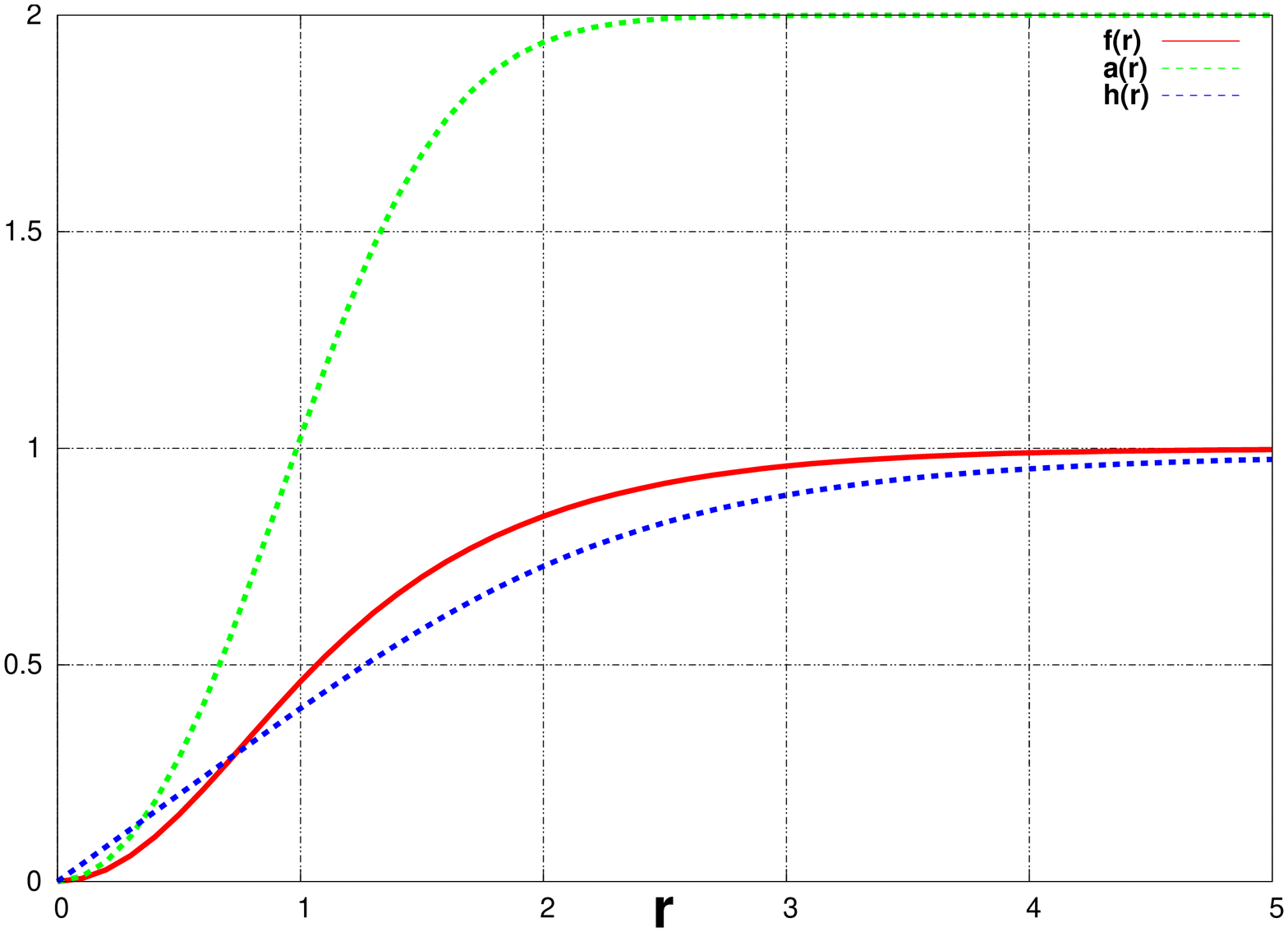} 
\end{center} 
\caption{Field profiles for different winding numbers.  
  The two upper panels show the field profiles for local-local 
  vortices with winding numbers $\left(1,2\right)$ on the left and 
  $\left(2,1\right)$ on the right. The lower panels  
  show the field profiles for local-global 
  vortices with winding numbers $\left(1,2\right)$ on the left and 
  $\left(2,1\right)$ on the right. 
  We denote by $f\left(r\right)$ the radial profile of the Higgs,  
  by $a\left(r\right)$ the  the radial profile of the gauge field  
  that couples to the Higgs, $h\left(r\right)$ for the axion 
  and finally $b\left(r\right)$ for the gauge field that couples to the 
  axion. For the local-global defect we simply set $b\left(r\right) = 0$
  along the entire radial direction. 
  \label{radial_field_profiles}} 
\end{figure}

One can then impose the appropriate boundary conditions 
depending on the winding number of phase of each scalar field.
The values of the fields at each point along 
the radial direction evolve according to 
Eq.(\ref{discretized_hamilton_eq}) and the final configuration 
determines the field profiles (see Fig.(\ref{radial_field_profiles})) 
and the corresponding energy of the topological defect.

\section{Lattice discretisation}
\label{app:lattice}
In order to carry out numerical simulations described in
Section~\ref{sec:results}, we discretised the equations of motion (\ref{equ:eom})
in the standard leap-frog fashion. 
The scalar fields $\phi$ and $\chi$ were defined at the lattice
sites, and their time derivatives $\dot\phi$ and $\dot\chi$ at 
temporal links between time
slices. We used temporal gauge $A_0=C_0=0$, and the spatial gauge fields $\vec{A}$ and $\vec{C}$ were represented by real numbers defined at links between lattice sites.

\def\x{{\vec{x}}}
\def\hi{{\hat\imath}}

In order to write the equations of motion in a compact form, we define the link variables
$U_{i}=\exp\left(-ie_1\delta x 
A_{i}\right)$ and $V_{i}=\exp\left(-ie_2\delta x 
C_{i}\right)$, 
and the forward and backward lattice derivatives of any function $f_{(\vec{x})}$,
\begin{equation}
\Delta^\pm_if_{(\vec{x})}=
\pm\delta x^{-1}\left(f_{(\vec{x}\pm\hi)}-f_{(\vec{x})}\right).
\end{equation}
We also define the lattice version of the covariant derivative
\begin{eqnarray}
D^+_i\phi_{(\vec{x})}&=&\delta x^{-1}\left(
U_{i,(\vec{x})}\phi_{(\vec{x}+\hi)}
-\phi_{(\vec{x})}\right),\nonumber\\
D^-_i\phi_{(\vec{x})}&=&\delta x^{-1}\left(
\phi_{(\vec{x})}-
U^*_{i,(\vec{x}-\hi)}\phi_{(\vec{x}-\hi)}\right)
,
\end{eqnarray}
with $U$ replaced by $V$ if the derivative is acting on $\chi$.

In this notation, the discretized equations of motion with the damping term $\sigma$
are
\begin{eqnarray}
\Delta_t \dot A_{i,(t,\x)}&=&
-\sigma \dot A_{i,(t-\delta t,\x)}
-\epsilon_{ijk}\epsilon_{klm}
\Delta^-_j\Delta^+_lA_{i,(t,\vec{x})}
-2e_1{\rm Im}\phi^*_{(t,\x)}D^+_i\phi_{(t,\x)}
,
\nonumber\\
\Delta_t \dot C_{i,(t,\x)}&=&
-\sigma \dot C_{i,(t-\delta t,\x)}
-\epsilon_{ijk}\epsilon_{klm}
\Delta^-_j\Delta^+_lC_{i,(t,\vec{x})}
-2e_2{\rm Im}\chi^*_{(t,\x)}D^+_i\chi_{(t,\x)}
,
\nonumber\\
\Delta_t\dot\phi_{(t,\x)}&=&
-\sigma \dot \phi_{(t-\delta t,\x)}
+D_i^-D_i^+\phi_{(t,\x)}
-\lambda_1\left(|\phi_{(t,\vec{x})}|^2-\eta_1^2\right)\phi_{(t,\vec{x})}
\nonumber\\
& & -\frac{\lambda_2}{2}\left(|\chi_{(t,\vec{x})}|^2-\eta_2^2\right)^2\phi_{(t,\vec{x})},
\nonumber\\
\Delta_t\dot\chi_{(t,\x)}&=&
-\sigma \dot \chi_{(t-\delta t,\x)}
+D_i^-D_i^+\chi_{(t,\x)}
-\lambda_2|\phi_{(t,\vec{x})}|^2\left(|\chi_{(t,\vec{x})}|^2-\eta_2^2\right)\chi_{(t,\vec{x})}
,\nonumber\\
\Delta_t A_{i,(t+\delta t,\x)}&=&
\dot A_{i,(t,\x)},
\nonumber\\
\Delta_t C_{i,(t+\delta t,\x)}&=&
\dot C_{i,(t,\x)},
\nonumber\\
\Delta_t\phi_{(t+\delta t,\x)}&=&\dot\phi_{(t,\x)},
\nonumber\\
\Delta_t\chi_{(t+\delta t,\x)}&=&\dot\chi_{(t,\x)},
\end{eqnarray}
where $\Delta_t\phi_{(t)}
=\delta t^{-1}[\phi_{(t)}-\phi_{(t-\delta t)}]$ etc.

In the temporal gauge, the fields also have to satisfy the Gauss laws,
\begin{eqnarray}
\label{equ:Gauss}
\sum_i\Delta^-_i\dot A_{i,(t,\x)}&=&
2e_1{\rm Im}\phi^*_{(t,\vec{x})}
\dot\phi_{(t,\vec{x})},\nonumber\\
\sum_i\Delta^-_i\dot C_{i,(t,\x)}&=&
2e_2{\rm Im}\chi^*_{(t,\vec{x})}
\dot\chi_{(t,\vec{x})},
\end{eqnarray}
which appear as additional constraints for the initial conditions.


\section*{References} 
\begin{thebibliography}{99} 
\bibitem{Spergel:2006hy} 
  D.~N.~Spergel {\it et al.}  [WMAP Collaboration], 
  arXiv:astro-ph/0603449. 
 
\bibitem{Guth:1980zm} A.~H.~Guth,  
  Phys.\  Rev.\ D {\bf 23}, 347 (1981).   
 
\bibitem{Jeannerot:2003qv} 
  R.~Jeannerot, J.~Rocher and M.~Sakellariadou, 
  Phys.\ Rev.\  D {\bf 68}, 103514 (2003) 
  [arXiv:hep-ph/0308134]; 
  M.~Sakellariadou, 
  arXiv:hep-th/0702003. 
 
\bibitem{Piran:1986dh} 
  T.~Piran, 
  Phys.\ Lett.\  B {\bf 181}, 238 (1986); 
  D.~S.~Goldwirth, 
  Phys.\ Rev.\  D {\bf 43}, 3204 (1991); 
  E.~Calzetta and M.~Sakellariadou, 
  Phys.\ Rev.\  D {\bf 45}, 2802 (1992); 
  E.~Calzetta and M.~Sakellariadou, 
  Phys.\ Rev.\  D {\bf 47}, 3184 (1993) 
  [arXiv:gr-qc/9209007]; 
  G.~W.~Gibbons and N.~Turok, 
  arXiv:hep-th/0609095; 
  C.~Germani, W.~Nelson and M.~Sakellariadou, 
  arXiv:gr-qc/0701172. 
 
\bibitem{Kallosh:2007ig} 
  R.~Kallosh, 
  arXiv:hep-th/0702059. 
 
\bibitem{Kachru:2003aw} 
  S.~Kachru, R.~Kallosh, A.~Linde and S.~P.~Trivedi, 
  Phys.\ Rev.\  D {\bf 68}, 046005 (2003) 
  [arXiv:hep-th/0301240]. 
 
\bibitem{Quevedo:2002xw} 
  F.~Quevedo, 
  Class.\ Quant.\ Grav.\  {\bf 19}, 5721 (2002) 
  [arXiv:hep-th/0210292]. 
 
\bibitem{Kachru:2003sx}  
  S.~Kachru, R.~Kallosh, A.~Linde, J.~M.~Maldacena, L.~McAllister  
  and S.~P.~Trivedi,  
  JCAP {\bf 0310}, 013 (2003) [arXiv:hep-th/0308055].   
 
\bibitem{Durrer:2005nz} 
  R.~Durrer, M.~Kunz and M.~Sakellariadou, 
  Phys.\ Lett.\  B {\bf 614}, 125 (2005) 
  [arXiv:hep-th/0501163]. 
 
\bibitem{Sarangi:2002yt} 
  S.~Sarangi and S.~H.~H.~Tye, 
  Phys.\ Lett.\  B {\bf 536}, 185 (2002) 
  [arXiv:hep-th/0204074]; 
  N.~T.~Jones, H.~Stoica and S.~H.~H.~Tye, 
  Phys.\ Lett.\  B {\bf 563}, 6 (2003) 
  [arXiv:hep-th/0303269]; 
 
\bibitem{Polchinski:2004hb} 
  J.~Polchinski, 
  AIP Conf.\ Proc.\  {\bf 743}, 331 (2005) 
  [Int.\ J.\ Mod.\ Phys.\  A {\bf 20}, 3413 (2005)] 
  [arXiv:hep-th/0410082]. 
 
\bibitem{Nielsen:1973cs}
  H.~B.~Nielsen and P.~Olesen,
  Nucl.\ Phys.\  B {\bf 61}, 45 (1973).
 
\bibitem{Jackson:2004zg} 
  M.~G.~Jackson, N.~T.~Jones and J.~Polchinski, 
  JHEP {\bf 0510}, 013 (2005) 
  [arXiv:hep-th/0405229]. 
 
\bibitem{Donaire:2005qm}
  M.~Donaire and A.~Rajantie,
  Phys.\ Rev.\  D {\bf 73}, 063517 (2006)
  [arXiv:hep-ph/0508272].
 
 
\bibitem{Sen:1997xi} 
  A.~Sen, 
  JHEP {\bf 9803}, 005 (1998) 
  [arXiv:hep-th/9711130]; 
  D.~Spergel and U.~L.~Pen, 
  Astrophys.\ J.\  {\bf 491}, L67 (1997) 
  [arXiv:astro-ph/9611198]; 
  M.~Bucher and D.~N.~Spergel, 
  Phys.\ Rev.\  D {\bf 60}, 043505 (1999) 
  [arXiv:astro-ph/9812022]. 
 
\bibitem{Sakellariadou:2004wq} 
  M.~Sakellariadou, 
  JCAP {\bf 0504}, 003 (2005) 
  [arXiv:hep-th/0410234]; 
 
\bibitem{Avgoustidis:2004zt} 
  A.~Avgoustidis and E.~P.~S.~Shellard, 
  Phys.\ Rev.\  D {\bf 71}, 123513 (2005) 
  [arXiv:hep-ph/0410349]; 
 
\bibitem{Tye:2005fn} 
  S.~H.~Tye, I.~Wasserman and M.~Wyman, 
  Phys.\ Rev.\  D {\bf 71}, 103508 (2005) 
  [Erratum-ibid.\  D {\bf 71}, 129906 (2005)] 
  [arXiv:astro-ph/0503506]; 
 
\bibitem{Copeland:2005cy} 
  E.~J.~Copeland and P.~M.~Saffin, 
  JHEP {\bf 0511}, 023 (2005) 
  [arXiv:hep-th/0505110]; 
 
\bibitem{Saffin:2005cs} 
  P.~M.~Saffin, 
  JHEP {\bf 0509}, 011 (2005) 
  [arXiv:hep-th/0506138]; 
 
\bibitem{Avgoustidis:2005nv} 
  A.~Avgoustidis and E.~P.~S.~Shellard, 
  Phys.\ Rev.\  D {\bf 73}, 041301 (2006) 
  [arXiv:astro-ph/0512582]; 

\bibitem{Hindmarsh:2006qn} 
  M.~Hindmarsh and P.~M.~Saffin, 
  JHEP {\bf 0608}, 066 (2006) 
  [arXiv:hep-th/0605014]. 
 
\bibitem{Avgoustidis:2007aa}
  A.~Avgoustidis and E.~P.~S.~Shellard,
  [arXiv:0705.3395].

\bibitem{Copeland:2006eh} 
  E.~J.~Copeland, T.~W.~B.~Kibble and D.~A.~Steer, 
  Phys.\ Rev.\ Lett.\  {\bf 97}, 021602 (2006) 
  [arXiv:hep-th/0601153].; 
  E.~J.~Copeland, T.~W.~B.~Kibble and D.~A.~Steer, 
  Phys.\ Rev.\  D {\bf 75}, 065024 (2007) 
  [arXiv:hep-th/0611243]. 
 
\bibitem{Johnson:2000ch} 
  C.~V.~Johnson, 
  arXiv:hep-th/0007170. 
 
\bibitem{Copeland:2003bj} 
 E.~J.~Copeland, R.~C.~Myers and J.~Polchinski, 
  JHEP {\bf 0406}, 013 (2004) 
  [arXiv:hep-th/0312067]; 
  J.~Polchinski, 
  arXiv:hep-th/0412244. 
 
\bibitem{Gubser:2004qj} S.~S.~Gubser, C.~P.~Herzog and I.~R.~Klebanov, 
  JHEP {\bf 0409}, 036 (2004) [arXiv:hep-th/0405282]. 
 
\bibitem{Herzog:2001fq} 
  C.~P.~Herzog and I.~R.~Klebanov, 
  Phys.\ Lett.\  B {\bf 526}, 388 (2002) 
  [arXiv:hep-th/0111078]. 
 
\bibitem{Klebanov:2000hb} 
  I.~R.~Klebanov and M.~J.~Strassler, 
  JHEP {\bf 0008}, 052 (2000) 
  [arXiv:hep-th/0007191]. 
 
\bibitem{Firouzjahi:2006vp} 
  H.~Firouzjahi, L.~Leblond and S.~H.~Henry Tye, 
  JHEP {\bf 0605}, 047 (2006) 
  [arXiv:hep-th/0603161]. 
 
\bibitem{Kibble:1976sj}
  T.~W.~B.~Kibble,
  J.\ Phys.\ A  {\bf 9}, 1387 (1976).
  
\bibitem{Hindmarsh:2000kd}
  M.~Hindmarsh and A.~Rajantie,
  Phys.\ Rev.\ Lett.\  {\bf 85}, 4660 (2000)
  [arXiv:cond-mat/0007361].

\bibitem{Moore:2006ec} 
 G.~D.~Moore and H.~Stoica, 
 Phys.\ Rev.\  D {\bf 74}, 065003 (2006) 
 [arXiv:hep-th/0605070]; 
 

\end{thebibliography}
\end{document}